\documentclass{aa}
\usepackage{natbib}
\bibpunct{(}{)}{;}{a}{}{,}
\usepackage{graphicx}
\usepackage{txfonts}
\usepackage[]{hyperref}
\usepackage{tikz}

\usepackage{amsmath,amssymb,latexsym}
\begin{document}

\title{Misidentification of Short GRBs as Magnetars in Nearby Galaxies}

\author{Elisa C. Sch\"osser
  \inst{1,2,3}\thanks{Fellow of the International Max Planck Research School for
    Astronomy and Cosmic Physics at the University of Heidelberg (IMPRS-HD)},
    J. Michael Burgess \inst{1},
  \and Jochen Greiner\inst{1} }

\institute{Max-Planck-Institut für Extraterrestrische Physik, Giessenbachstrasse
  1, 85748 Garching, Germany,\\ \email{jburgess@mpe.mpg.de} \and Technische
  Universit\"at M\"unchen, Physik-Department, James-Frank-Str. 1, 85748 Garching
  bei M\"unchen, Germany 
  \and Zentrum f\"ur Astronomie der Universit\"at Heidelberg, Astronomisches
  Rechen-Institut, M\"onchhofstr. 12-14, 69120 Heidelberg, Germany,
  \email{elisa.schoesser@uni-heidelberg.de}}

\date{}

\abstract { Recent observations of GRB 200415A, a short and very bright pulse of
  $\gamma$-rays, have been claimed to be an extragalactic magnetar giant flare
  (MGF) whose proposed host galaxy is the nearby ${\mathrm{NGC} \,
    253}$. However, as the redshift of the transient object was not measured, it
  is possible that the measured location of the transient on the celestial
  sphere and the location of the local galaxy merely coincided. Thus, its real
  progenitor could have been arbitrarily far away, leading possibly to a much
  larger luminosity of the transient, and leaving the standard model of short
  gamma-ray bursts (sGRBs), the merger of two compact objects, as an explanation
  for the observations.}
{In this study, our aim is to compute the false-alarm rate for the
  misinterpretation of sGRBs as magnetars in a given observation period.}
{We simulate synthetic surveys of sGRB observations in a time period of 14 years
  corresponding to the operation period of the Gamma-ray Burst Monitor (GBM)
  detector. For all sGRBs that align on the sky with a nearby Local Volume
  galaxy, we generate realistic data which is folded through the response of the
  GBM. To identify candidates of sGRBs that may be misinterpreted as magnetars,
  six selections (spatial, star formation rate, GBM trigger, duration, isotropic
  energy release, and fluence) are applied to the simulated surveys.}
{In a non-negligible fraction, 15.7 \%, of the simulated surveys, we identify at
  least one sGRB that has the same characteristics as a magnetar giant flare and
  could be thus misinterpreted as magnetar. Thus, we conclude that the
  selections that were proposed in previous work to unambiguously identify an
  extragalactic magnetar giant flare are not sufficient.}
{}


\maketitle

\section{Introduction}
Time-domain astronomy is a rapidly growing field with ever more instruments
dedicated to measuring transient events in the sky. As the number of events
grow, the likelihood of chance coincidence between these events or with
temporally static astrophysical objects also grows. In this young field, the
proper statistical techniques to compute these chance coincident detection
probabilities do not fully exist. Frustrating the problem further is that often
there are not physical models for the source rates, cosmological distribution,
and/or source production processes which would allow for one to rule out chance
association with purely astrophysical arguments. When the events are rare or
even from unexplained origins, the problem intensifies. Herein, we examine one
such association where a transient gamma-ray event spatial coincided with a
nearby galaxy in attempt to determine the probability that this association was
misidentified.

There are two known astrophysical sources that emit short, transient pulses of
$\gamma$-rays, namely magnetars and short gamma-ray bursts (GRBs).  With
luminosities of up to $10^{54}$ erg s$^{-1}$ \citep{2013Frederiks}, GRBs are the
most luminous explosions in the electromagnetic spectrum known in the
universe. The prompt emission of GRBs peaks in the $\gamma$-ray band, and is
followed by an afterglow that may be detectable in the X-ray, optical and radio
wavelengths, if the follow-up observations are fast and sensitive enough.
Typically, GRBs are classified by their duration into short ($< 2$ s) and long
GRBs ($>2$ s) \citep{Mazets1981}. Short GRBs (sGRBs) are associated with the
merger of two compact objects, which was confirmed by the joint detection of a
short GRB and a gravitational wave (${\mathrm{GW \, 170817}}$) in
2017\citep{Abbot2017}.

First indicated by their isotropic spatial distribution on the sky and later
confirmed by redshift measurements, it was shown that GRBs originate at
cosmological distances \citep{1997Metzger}. From absorption lines in the optical
afterglow spectrum, the redshift can be directly determined. However, this is
not always possible, as the intensity of the optical afterglow decays rapidly
with time, and is often lower than the detection limit at the time of follow-up
observations. Thus, the most frequently used method for determining the redshift
is to attempt to identify a host galaxy of the GRB and measure its
distance. Only about 24 \% of all GRBs have determined
redshifts\footnote{Statistic
  from \url{https://www.mpe.mpg.de/~jcg/grbgen.html}}. The search for host
galaxies in the field of view can lead to wrong associations, especially if the
GRB's localization error is larger than a few arcseconds. The existence of a
bright galaxy in the field of view that potentially has a large angular size may
prevent the detection of the dimmer true host of the GRB. One recent example for
an initial wrong association of a GRB with a galaxy that was later discarded by
deeper follow-up observations was the detection of {GRB 220611A}. First, the GRB
triggered the Swift Burst Alert Telescope \citep{2022Cenko} and the GRB was
localized with an uncertainty of 3 arcmin. A nearby galaxy with redshift
$z=0.049$ was proposed as a candidate for the host galaxy as it had an angular
distance of 15 arcsec from the GRB. Later, \citet{2022Rastinejad} found another
faint optical source in the Legacy Survey \citep{2019Dey} that coincided with
the location of the detected GRB. A follow-up observation with Chandra lead to
the detection of the GRB's X-ray afterglow and decreased the uncertainty of the
localization to only 0.5 arcsec \citep{2022Levan}. At exactly that 0.5 arcsec
transient location, the host galaxy was identified, and spectroscopically
determined to be at z=2.36 \citep{2022Schneider}. Prior to this identification
the association to the much brighter, nearby galaxy was (wrongly) preferred.

Magnetars, on the other hand, are believed to be young and strongly magnetized
neutron stars with a magnetic field strength of $10^{13}-10^{15}$ G
\citep{1992Duncan}. Magnetic stresses at the core-crust boundary or
instabilities in the magnetic field can cause the sudden ejection of magnetic
energy which can be observed as repeating short bursts in the X-rays and soft
$\gamma$-rays \citep{1993Thompson}. Magnetars are mainly found within the Milky
Way and were discovered by the detection of recurrent short bursts in the X-rays
and soft gamma-rays with luminosities of about $10^{40}$ erg s$^{-1}$.
Observations of flares with an energy output of more than $10^{44}$ erg s$^{-1}$, called magnetar giant flares (MGFs), were observed for three already known
magnetars after an increase in activity, {SGR 0526-66}
\citep{1980Evans,1979Mazetsa}, {SGR 1900+14} \citep{1999Hurley,1999Mazets}, and
{SGR 1806-20} \citep{2005Palmer,2005Mereghetti}. These three magnetars are of
galactic origin, and were characterized by an initial strong pulse followed by a
long tail with quasi-periodic oscillations (QPOs) in the frequencies between 18
and 626.5 Hz \citep{1983Barat,2005Israel,2005Strohmayer,2006Watts,2018Pumpe}.
Recurrent MGFs have not been observed yet.\\
The total released energy of $(2-3)\times10^{46}$ erg of the SGR 1806-20 giant
flare suggested that MGFs may be detectable up to distances of about 50 Mpc,
including other nearby galaxies as origin \citep{2005Hurley}. This finding
motivated searches in measured data of sGRBs to identify a fraction of GRBs that
can be associated with nearby galaxies and can thus be explained by MGFs
\citep{2005Tanvir,2010Tikhomirova,2015Svinkin,2018Mandhai}. Following this
campaign, additional extragalactic MGF candidates were proposed, namely {GRB
  051103} \citep{Ofek2006}, {GRB 070201} \citep{Mazets2008, Ofek2008}, {GRB
  070222} \citep{Burns2021}, and most recently {GRB 200415A}
\citep{Svinkin2021,Roberts2021}. The main argument for the interpretation of the
observations as magnetar giant flares was the coinciding locations of the
detected $\gamma$-rays with nearby galaxies that have a distance smaller than 5
Mpc. {GRB 051103} was found to coincide with the location of the M81/M82 galaxy
group, {GRB 070201} was associated with the Andromeda Galaxy, {GRB 070222} with
M83, and {GRB 200415A} with {NGC 253}. However, no distance estimate was
possible for any of the proposed extragalactic MGFs. Thus, it cannot be excluded
that the origin of the gamma-rays is at much farther distances beyond the
associated galaxies. As the measured flux scales approximately as
$\propto 1/r^2$, a larger distance $r$ would correspond to a higher luminosity
of the source. Beyond a certain distance (1-2 Mpc), the origin of gamma-rays
from magnetars becomes unlikely, leaving sGRBs as natural explanation. Thus, a
distance estimate is crucial for the distinction of magnetar vs. sGRB
interpretation.

Given that all claimed extragalactic magnetars observations lack a distance
estimate, the aim of this work is to compute the probability for the
misinterpretation of an sGRB as a magnetar due to chance alignment with a nearby
galaxy.  To do so, we generate a population of sGRBs as seen by Fermi-GBM via
sampling all population parameters from observationally motivated
distributions. We select the sGRBs that are spatially coincident with nearby
galaxies. For the chosen sGRBs, we simulate realistic GBM data, analyze the
simulated data, and select candidates for misinterpretation as magnetars based
on multiple selection criteria that were applied in the past to identify
extragalactic MGFs.

The article is structured as follows: In Sec. \ref{sec:popsynth}, we describe
the generative model used for the population synthesis. In the subsequent Sec.
\ref{sec:cosmogrb}, we explain how we simulate Fermi-GBM data. How we analyze 
the simulated data is summarized in Sec. \ref{sec:threeml}. We summarize the 
selection criteria in Sec. \ref{sec:selection_criteria}, which we apply to our 
sGRB surveys to identify candidates for misinterpretation as extragalactic 
MGFs. To proof the consistency of our simulations, we performed tests which are 
presented in Sec. \ref{sec:model_checks}. Finally, we show our results in Sec. 
\ref{sec:results}, and summarize our conclusions in Sec. \ref{sec:conclusions}.

\section{SGRB population synthesis}\label{sec:popsynth}
To simulate a population of sGRBs, we sample the latent parameters from
distributions which are motivated by observations. The latent parameters of 
sGRBs include their sky coordinates, luminosity, distance, duration, and the
parameters describing their spectral evolution. The right ascension and
declination of the simulated GRBs is drawn in such a way that they are 
uniformly distributed on the sphere. In the following, we describe 
our chosen distributions for the other parameters.

\subsection{Duration, light curve and spectral shape}
In all simulations, we model the sGRBs with a constant light curve over the
burst duration. We note that this simplifies the realistic pulse profile of real
sGRBs but is necessitated by computation costs. The inclusion of realistic
profiles does not change our results.

\begin{figure}[htpb]
  \centering
  \includegraphics[width=0.49\textwidth]{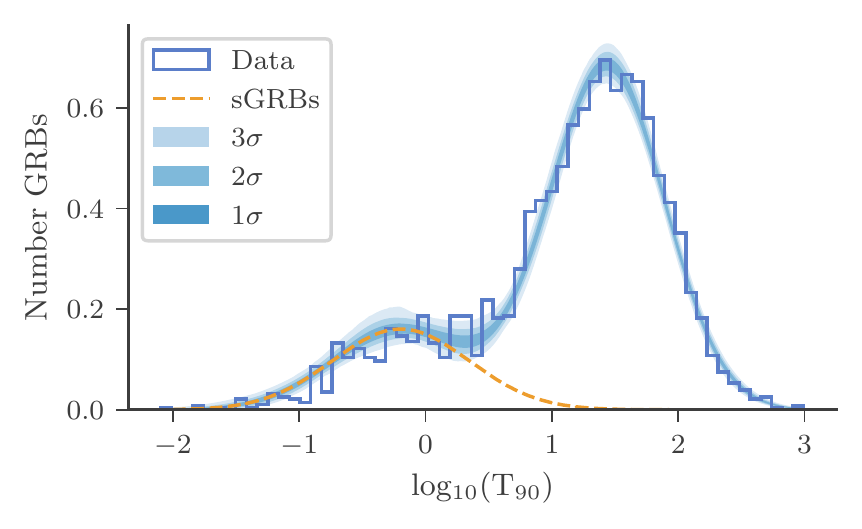}
  \caption{Histogram of the measured $\log_{10}(T_{90})$ data from GBM. Blue
    shaded: $68.3$ \% ($1\sigma$), $95.5$ \% ($2\sigma$), and $99.7$ \%
    ($3\sigma$) quantiles of the posterior $T_{90}$ distribution of the mixture
    model. Orange dashed line: Best fit normal distribution for the durations of
    sGRBs using the medians of the parameters' marginal posterior
    distributions.}
  \label{fig:t90_posterior}
\end{figure}

The measured duration ($T_{90}$) distribution is fitted with a mixture model of
two log-normal distributions. We use the measured durations of all 3291 GRBs in
the public Fermi/ GBM catalog
\citep{2020Kienlin}.  For the fit, we use the Bayesian framework Stan
\citep{Stan}. The uncertainty of the posterior fit is depicted in Figure
\ref{fig:t90_posterior}. The chosen median best fit distribution function for
sGRBs that we use in this work is highlighted in orange.

\begin{table}[tb]
  \centering
  \caption{Medians, modes, means, and 90 \% credible intervals (C.I.) of the
    marginal posterior distributions of the $T_{90}$ fit for the means 
    ($\mu_1$, $\mu_2$), standard deviations ($\sigma_1$, $\sigma_2$) and the
    mixture parameter ($\theta$).}
  \begin{tabular}{|c|c|c|c|c|}
    \hline
    Parameter   & Median    & Mode   & Mean   & 90 \% C.I. \\ \hline \hline
    $\mu_1$     &  -0.20  &  -0.20 &  -0.19 & (-0.29,-0.07)\\
    $\mu_2$     &  1.44   &  1.43  &  1.40  & (1.40,1.50)\\
    $\sigma_1$  &  0.54   &  0.55  &  0.55  & (0.46,0.65)\\
    $\sigma_2$  &  0.46   &  0.46  &  0.46  & (0.44,0.47)\\
    $\theta$    &  0.22   &  0.20  &  0.20  & (0.20,0.30)\\
    \hline
  \end{tabular}
  \label{tab:T90_best_fit}
\end{table}

In Table \ref{tab:T90_best_fit}, the median, mode, mean, and 90 \% credible
intervals (C.I.) of all fit parameters are listed. The best-fit parameters are
in similar range as the previous fit results in \citet{2020Kienlin}. Only the
result for the standard deviation differs approximately by a factor of two for
both short and long GRBs. The median best fit mixing parameter of 0.22 yields
that 22 \% of the GRBs in the sample are sGRBs, corresponding to 724 GRBs.

In this study, we draw the duration $T_{90}$ of the sGRBs from a
$\log_{10}$-normal distribution with a probability density function that is
given by
\begin{equation}
  p(T_{90}) = \frac{1}{\sqrt{2\pi} \,\sigma_1 \,T_{90}\, \ln(10)} \exp{-\frac{1}{2}\left(\frac{\log_{10}(T_{90})-\mu_1}{\sigma_1}\right)^2},
  \label{eq:log10norm_t90}
\end{equation}
where we use for the mean $\mu_1$ and standard deviation $\sigma_1$, the median
best-fit parameters as summarized in Table \ref{tab:T90_best_fit}.

It was shown that the largest fraction (69 \%) of GRB prompt emission spectra
are best fit by a cutoff power law (CPL) when compared to three other empirical
models, including the Band function (9.3 \%), a smoothly broken power law (11.4
\%), and a simple power law (10.2 \%) \citep{Yu2016}.  Thus, we model the
spectra of the sGRBs with CPLs. The CPL function is parameterized as
\begin{align}
  N(E;K,E_\mathrm{p},\alpha) = K
  \left(\frac{E}{E_\mathrm{piv}}\right)^{\alpha} \exp \left(-\frac{(2+\alpha) E}{E_\mathrm{p}}\right).
  \label{eq:cpl}
\end{align}
Here, $K$ is the normalization flux given in {cm$^{-2}$ s$^{-1}$ keV} at the
pivot energy $E_\mathrm{piv}$, and $\alpha$ is the spectral index at low
energies. In the $\nu F_\nu$ representation, the spectrum has its peak at
$E_\mathrm{p}$ which is given in keV.

The distribution for $E_\mathrm{p}$ is chosen to be the same as the best-fit
distribution in \citet{Ghirlanda2016} which is a broken power law function. The
parameters of the distribution are summarized in Table
\ref{tab:ghirlanda_fit_params}. Instead of fixing the power law index $\alpha$
as in \citet{Ghirlanda2016}, $\alpha$ is drawn from a truncated normal
distribution with a mean of $\mu_\alpha=-0.6$ and a standard deviation
$\sigma_\alpha = 0.2$ in the interval $[-1.5,0]$. The chosen distribution for
$\alpha$ is motivated by the results from \citet{2019Burgess}.

\subsection{Redshift and Luminosity Distribution}
To determine the redshift and luminosity distribution from observations is a
non-trivial task as the observed flux is both dependent on the luminosity and
redshift. Furthermore, as a result of various selection effects, the sample of
detected GRBs is not representative of the underlying population. One of the
selection effects is the so-called Malmquist bias which describes the effect of
preferably detecting bright sources. In comparison to long GRBs, it is even more
challenging to determine the redshift and luminosity distribution for short GRBs
as there is only a very small number of sGRBs with measured redshift, thus
enhancing the selection effects for sGRBs immensely. Nevertheless, there were
multiple attempts in the past to determine the redshift and luminosity function
of sGRBs by fitting the measured peak flux distribution and redshifts
\citep[e.g.][]{2005Guetta,2006Guetta,2006Nakar,2011Virgili,2014D'Avanzo,2015Wanderman}.
With the aim of decreasing selection effects and obtaining a representative
sample of GRBs, often a flux limit is set when choosing the GRBs. In addition to
using the peak flux distribution, \citet{Ghirlanda2016} include measurements of
the peak energy of the observed spectra, the fluence, the duration, and if the
redshift is known, the isotropic equivalent luminosity and energy, for the fits.

In this work, we sample the luminosity $L$ and redshift $z$ from the fitted
distributions $\Phi(L)$ and $\rho (z)$ in \citet{Ghirlanda2016}. Assuming that
the bursting rate of GRBs is proportional to the star formation rate, the
following parameterization of the redshift distribution
\begin{equation}
  \rho (z) = \dot{\rho_0} \Delta t_{\mathrm{obs}} \frac{1+p_1 z}{1 + (z/z_p)^{p_2}},
  \label{eq:redshift_distribution}
\end{equation}
is used, as in \cite{Cole2001}. The parameter $\dot{\rho_0}$ is the rate of GRBs
at redshift zero and has the units of yr$^{-1}$ Gpc$^{-3}$, and
$\Delta t_{\mathrm{obs}}$ is the observation period of the survey. The rise,
location of the peak, and decay of the redshift distribution are all influenced
by the parameters $z_p$, $p_1$, and $p_2$. The chosen redshift distribution is
independent of the emission properties of the GRBs, which implies the underlying
assumption that the emission properties do not change with cosmic time.

For the peak energy $E_\mathrm{p}$, a broken power law distribution was chosen
which has the probability density function
\begin{align}
  p(E_\mathrm{p}) \propto
  \begin{cases}
    \left(E_\mathrm{p}/E_{p,\mathrm{break}}\right)^{-a} \quad E_\mathrm{p} \leq E_{p,\mathrm{break}}\\
    \left(E_\mathrm{p}/E_{p,\mathrm{break}}\right)^{-b} \quad E_\mathrm{p} > E_{p,\mathrm{break}}
  \end{cases}
\end{align}
where $E_{p,\mathrm{break}}$ is the break energy at which the power law index
changes from $-a$ to $-b$.

We use the results of case (c) in \citet{Ghirlanda2016}, in which no correlation
between the peak energy $E_\mathrm{p}$ and the isotropic equivalent energy
$L_{\mathrm{iso}}$ is assumed. Thus, we draw the luminosity from the fitted
independent distribution, a broken power law distribution function
\begin{equation}
  \Phi(L) \propto \begin{cases}
                    (L/L_\mathrm{b})^{-\alpha_1} \quad L \leq L_\mathrm{b}, \\
                    (L/L_\mathrm{b})^{-a_2} \quad L > L_\mathrm{b}.
                  \end{cases}
                  \label{eq:bpl_lum}
                \end{equation}

\begin{table*}[t]
  \centering
  \caption[]{Modes of all fitted parameters in case (c) from the Monte Carlo
    Markov Chain results from \citet{Ghirlanda2016}.}
  \begin{tabular}{|l | c c c c c c c c c c|}
    \hline
    Parameter  & $\dot{\rho_0}$ [yr$^{-1}$ Gpc$^{-3}$]  &
                                                          $p_1$ & $p_2$ & $z_p$ & $a$ & $b$ & $E_{\mathrm{p}_b}$ [keV] &
                                                                                                                         $\alpha_1$ & $\alpha_2$ & $L_b$
                                                                                                                                                   $\left[10^{52} \, \mathrm{erg} \, \mathrm{s}^{-1}\right]$ \rule{0pt}{3ex} \rule[-2ex]{0pt}{0pt} \\  \hline\hline
    Mode (c)   &  0.8 & 2.0 & 2.0 & 2.8 & -0.55 & 2.5 & 2100 & -0.32 & 1.8 & 0.79 \rule{0pt}{2ex} \rule[-1ex]{0pt}{0pt}\\  \hline
  \end{tabular}
  \label{tab:ghirlanda_fit_params}
\end{table*}

The modes of the best-fit parameters from \citet{Ghirlanda2016} for the
redshift, peak energy, and luminosity distribution that we use in this work are
summarized in Table \ref{tab:ghirlanda_fit_params}.

\subsection{Number of objects in Simulated sGRB Surveys}
The total number of GRBs $N$ within the volume with maximum distance
$z_{\mathrm{max}}$ that is expected to be observed during the observation period
of $\Delta t_{\mathrm{obs}}$ is computed by
\begin{align}
  N &= \int\limits_{0}^{z_{\mathrm{max}}}  \frac{c \, d_L^2 \, \dot{\rho}'(z)}{(1+z)^3 \, H_0\sqrt{ \Omega_\mathrm{m}(1+z)^3 + \Omega_\Lambda }} \, \mathrm{d} z \int \mathrm{d}\Omega \int\limits_{\Delta t_{\mathrm{obs}}}\mathrm{d} t, \\
    &= \Delta t_{\mathrm{obs}} \, D \, \int\limits_{0}^{z_{\mathrm{max}}}  \frac{c \, d_L^2 \, \dot{\rho}'(z)}{(1+z)^3 \, H_0\sqrt{ \Omega_\mathrm{m}(1+z)^3 + \Omega_\Lambda }} \,  \mathrm{d}z,
      \label{eq:volumeintegral}
\end{align}
where $D$ is the duty cycle of the telescope. The duty cycle of GBM is
approximately 60 \% \citep{FermiWebpage} which describes the times per day at
which the instrument is shut down when flying through the South Atlantic Anomaly
(SAA), and due to the fraction of the sky which is occulted by Earth. For all
simulations of a survey of sGRBs, a total observing period of
$\Delta t_{\mathrm{obs}} = 14 \, \mathrm{years}$ was chosen, corresponding
approximately to the time since GBM was launched in 2008.

\subsection{Summary of Simulated Parameters and their Distributions}

\begin{figure*}[htpb]
  \centering
  \includegraphics[width=0.7\textwidth]{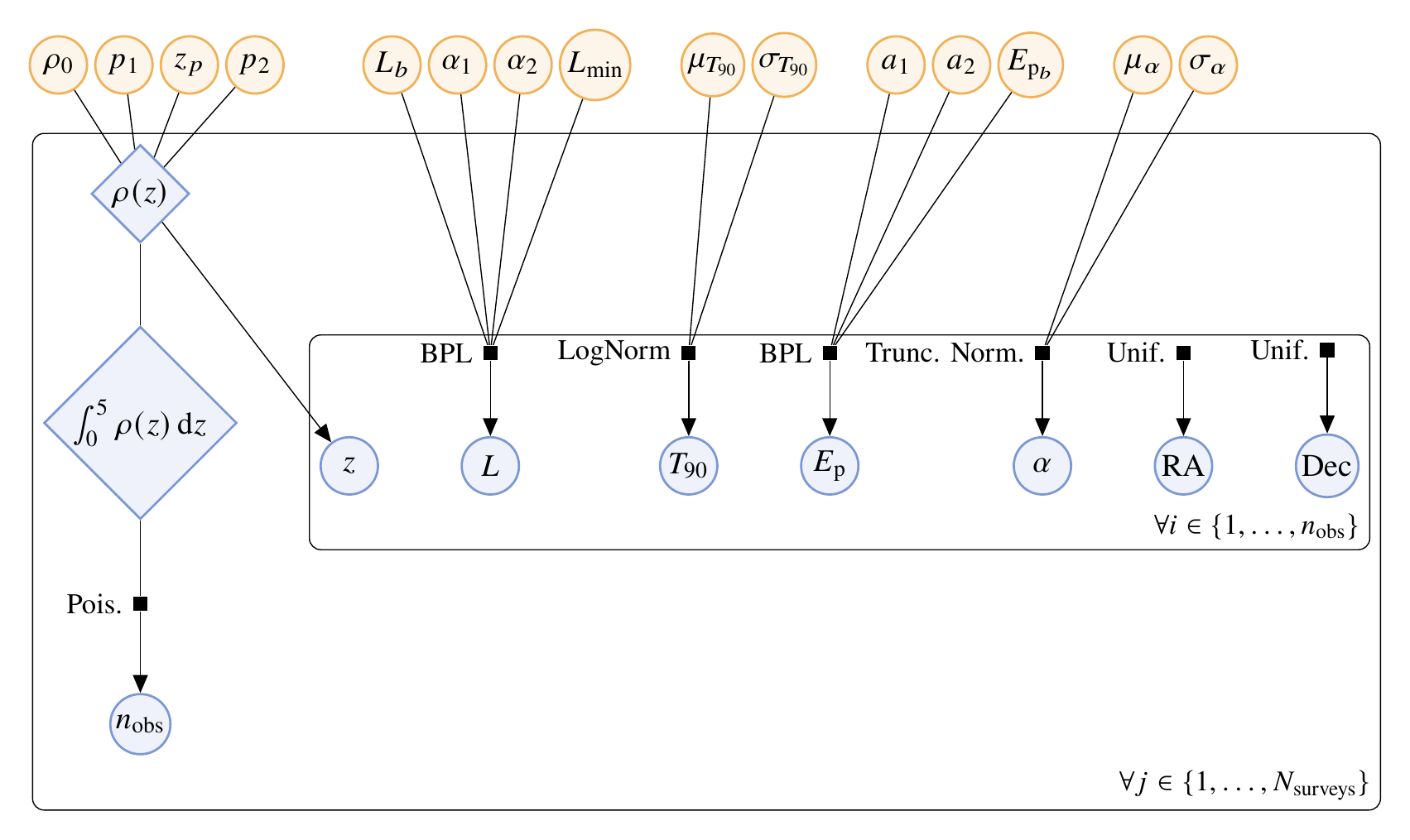}\\
  \caption{Probabilsitic graphical model summarizing all sampled latent
    parameters (\textbf{blue} circles) of each GRB with a constant temporal
    profile using the parameters defined in case (c) of \citet{Ghirlanda2016}
    for the luminosity $L$, redshift $z$, and peak energy $E_{\mathrm{p}}$
    distribution. The hyperparameters defining the used probability density
    distributions are highlighted with \textbf{orange} circles. }
  \label{fig:networks}
\end{figure*}

In this work, we use the object-oriented astrophysics population synthesis
framework \texttt{popsynth}
\footnote{\url{https://github.com/grburgess/popsynth}} \citep{Burgess2021}.
Once the population is set up and all parameter distributions and dependencies
are chosen, the expected total number of objects is computed by integrating the
spatial rate distribution. The number of detected GRBs in the survey is then
drawn from a Poisson distribution and for all objects in the survey, the latent
and observed quantities defining the population are drawn, including the
distances and luminosities of the objects. The final population object
containing all sampled parameters is saved as an HDF5 file for later analysis.
In Fig. \ref{fig:networks}, all sampled parameters, and their distributions
(with needed parameters) are summarized.

\section{Simulating GBM Data}\label{sec:cosmogrb}

Once the population of objects is created, we simulate Fermi-GBM data so that we
can analyze its properties and have realistic event triggers. In order to
obtain realistic GBM data for GRBs in the simulated sample, we designed a
generic software package
\texttt{cosmogrb}\footnote{\url{https://github.com/grburgess/cosmogrb}} which
accepts as input a population of latent parameters generated with
\texttt{popsynth} and outputs a corresponding catalog of synthetic GBM
Time-tagged event (TTE) data for all detectors for each GRB. These data can then
be analyzed as if they were an authentic GRB observation both to determine if
the event would have resulted in a GBM trigger alert as well as for routine
spectral analysis. The steps of the simulation are detailed below.

\subsection{Orbit and Detector Response}
As the orientation of the various GBM detectors with respect to a transient
event is crucial to both the detection and observed spectral data properties of
the event, we must simulate how the detectors' orientation changes with respect
to the celestial sky as GBM orbits the Earth. To achieve this, \texttt{cosmogrb}
takes as input the so-called position history file which details both the
orbital location of the Fermi spacecraft as well as its zenith pointing. Thus,
we fully account for the rocking motion performed by Fermi so that the Large
Area Telescope (LAT) can observe the full sky. For a given simulated GRB event,
a random point within the orbit is selected. With the orientation of the GBM
detectors set for the observation, this orientation and the celestial location
of the simulated GRB are passed to
\texttt{gbmdrmgen}\footnote{\url{https://github.com/grburgess/gbm_drm_gen}}
which is used to generate the correct detector responses for all GBM detectors
corresponding to the event.

\subsection{Spectrum Simulation}

With the detector responses defined for the observation, the latent source
spectrum can be folded through these responses to create the observed counts as
the spectrum evolves with time. In order to create these counts, the spectrum as
a function of time and energy is multiplied by a GBM detector's effective area
as a function of energy and then integrated over energy to compute the total
photon flux as a function of time. This is then used to sample the exponentially
distributed arrival times of the observed counts. This sampling is done via a
rejection scheme for an inhomogeneous Poisson process as described in
\citet{2021Burgess}. Once the arrival times have been selected, they are passed
to the effective area weighted source spectrum as the given time which is then
rejection sampled over energy to obtain the energy of the observed photon. This
energy is then redshiftted and passed to the energy redistribution function
corresponding to its value to compute the sampled PHA channel of the count. The
end result are time-tagged counts corresponding to PHA channels which can then
be converted to the standard GBM TTE data format for each detector.

In order to add a background to each observation in each detector,
\texttt{cosmogrb} uses a set of template background count spectra which have
their amplitudes randomized to reflect the observed variations in the GBM
background. The backgrounds are assumed to be constant in time and have their
count arrival times sampled as described above. The PHA channels of the
simulated counts are then sampled after being weighted with the distributions
from the templates.

\section{Analysis of GBM Data}\label{sec:threeml}
We analyze the simulated data using the same framework as previously detected
GRBs in the past using the Multi-Mission Maximum Likelihood (\texttt{3ML})
framework \citep{threeml2015}
\footnote{\url{https://github.com/threeML/threeML}}. Our spectral analysis is
based on the same methodology as presented in \citet{2019Burgess}.

For all simulated GRBs that were selected by the GBM trigger, we analyze the
simulated data. We use the Bayesian block algorithm \citep{2013Scargle} to find
significant changes in the count rates. If less than three time bins are found
by the algorithm, and thus no significant flux increase is found, we stop the
GRB analysis, and we discard the GRB from the sample.

If three or more time intervals are found, we fit the background in the first
and last time interval with a constant by maximizing the Poisson likelihood that
is given by
\begin{equation}
  \mathcal{L} = \prod_{i=1}^{N} \frac{M^{B_i}e^{-M}}{B_i!},
\end{equation}
where $N$ is the number of time bins, $M$ is the number of model counts given by
the constant background rate, and $B_i$ is the number of measured background
counts \citep{2016Greiner}.

We use profile likelihoods in which the likelihood is only evaluated at the
values for $\vec{\alpha_b}$ that minimize the negative log-likelihood function
for specific constant source parameters $\vec{\alpha_s}$
\begin{equation}
  - \ln \mathcal{L}(\alpha_s) = \min_{\vec{\alpha_b}}[-\ln(\mathcal{L}(\vec{\alpha_s}, \vec{\alpha_b}))],
  \label{eq:profile_likelihood}
\end{equation}
as described in \citet{threeml2015}. After the background model $\widetilde{B}$
is fitted in the off-source time period, it has Gaussian uncertainties with
standard deviation $\sigma_{\widetilde{B}}$. In contrast, the total counts are
still Poisson distributed. Thus, a Poisson-Gaussian likelihood is needed to fit
the spectral model to the data. To do so, we use the so-called PG-statistic\footnote{Definition is taken
  from
  \url{https://heasarc.gsfc.nasa.gov/xanadu/xspec/manual/XSappendix Statistics.html.}}
\begin{equation}
  -2 \ln \mathcal{L} = 2 \sum_{i=1}^{N} M_i + t_s f_i - S_i \ln(M_i t_s f_i) + \frac{(\widetilde{B}_i - t_s f_i)^2}{2\sigma_{\widetilde{B},i}^2} - S_i (1- \ln S_i)
\end{equation}
for each $i^{\mathrm{th}}$ energy bin. Here, $N$ is the number of total counts,
$S_i$ is the number of measured source counts, $M_i$ is the predicted number of
counts based on the model and the response, $t_s$ is the source time interval,
and $f_i$ is the profiled-out background model which is derived by setting the
derivative of the likelihood as in Equation \eqref{eq:profile_likelihood} to
zero. The spectrum is rebinned in such a way that there is at least one count
per bin, as this is required for using profile likelihoods. We use nested
sampling (\texttt{MultiNest} \citep{multinest2009, pymultinest2014}) for
computing the posterior distributions.

Once, the time-integrated spectrum $N(E;K,E_\mathrm{p},\alpha)$ is fitted, we
can compute the total isotropic energy release over the standard $1-10^4$ keV
range for a given source time interval $\Delta t$ by integrating the observed
time-integrated spectrum over energy
\begin{equation}
  E_{\mathrm{iso,\Delta t}} = 4\pi d_L^2 \int_{1 \mathrm{\, keV}}^{10^4 \mathrm{\,keV}}
  E \, N(E;K,E_\mathrm{p},\alpha) \, \mathrm{d} E.
  \label{eq:Eiso}
\end{equation}
By summing the computed isotropic energy from all found source time intervals,
the total isotropic energy release can be determined.

\section{Selection Criteria}\label{sec:selection_criteria}
For the search of MGF candidates in the simulated universes of GRBs, we
implemented multiple selections that are motivated by the selection criteria
that were used to identify extragalactic MGFs in the past. The selection
criteria are summarized in the following.

\subsection{Spatial Selection}
The first selection is applied to the sampled GRB coordinates. We only select
GRBs that can be associated on the sky with a nearby galaxy.  We use the Local
Volume (LV) Galaxy catalog of
\citet{2012Kaisina}\footnote{\url{https://www.sao.ru/lv/lvgdb}} which currently
contains information about 1246 galaxies up to a distance of 11 Mpc from the
Milky Way.  The catalog is an extension of the Catalog of Neighboring Galaxies
\citep{2004Karachentsev}.

\begin{figure}[htbp]
  \centering
  \includegraphics[width=0.49\textwidth]{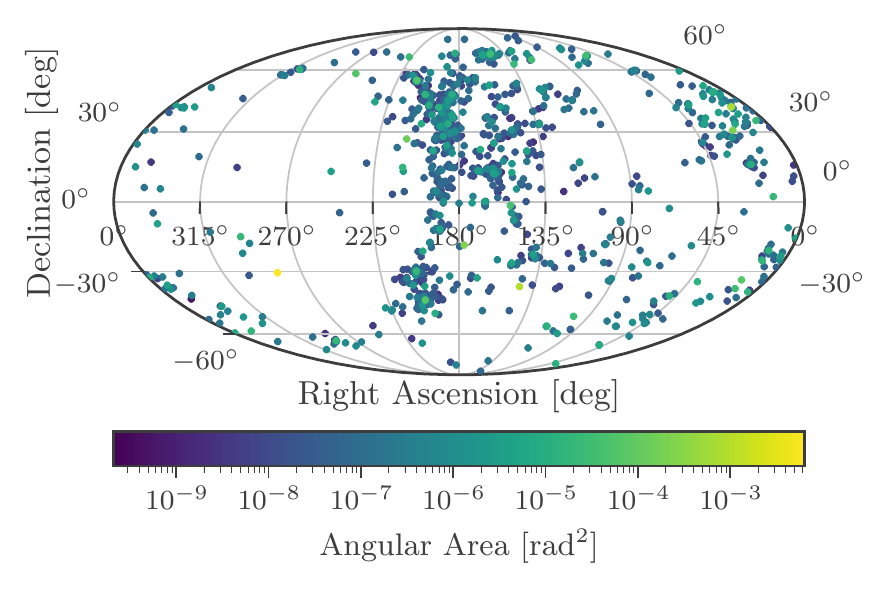}
  \caption{Location of the Local Volume galaxies on the sky in ICRS coordinates
    as given in the catalog of \citet{2012Kaisina}. The color illustrates the
    angular extent of the galaxies.}
  \label{fig:lv_gal_loc}
\end{figure}

For all LV galaxies, the center of the galaxy on the sky is given by its right
ascension $\phi_{\mathrm{cen}}$, and declination $\theta_{\mathrm{cen}}$. We
approximate the form of a galaxy as an ellipse which is projected onto a sphere
with distance $d$, with semi-major and semi-minor angular axis, $\alpha$ and
$\beta$ respectively. The angular size of the elliptical galaxies is then given
by $ {A = \pi \, \alpha \, \beta}$. In the sky map in Fig. \ref{fig:lv_gal_loc},
the location of all galaxies and their angular size are shown. For the
coordinates, we use the International Celestial Reference Frame (ICRS). As the
Large and Small Magellanic Cloud would dominate the results with their
exceptional large angular sizes of $A_{\mathrm{LMC}} = 8.6 \times 10^{-2}$
rad$^2$ and $A_{\mathrm{SMC}}= 2.2 \times 10^{-2}$ rad$^2$, we excluded them
from our analysis.

\begin{figure*}[htbp]
  \centering
  \includegraphics[width=0.49\textwidth]{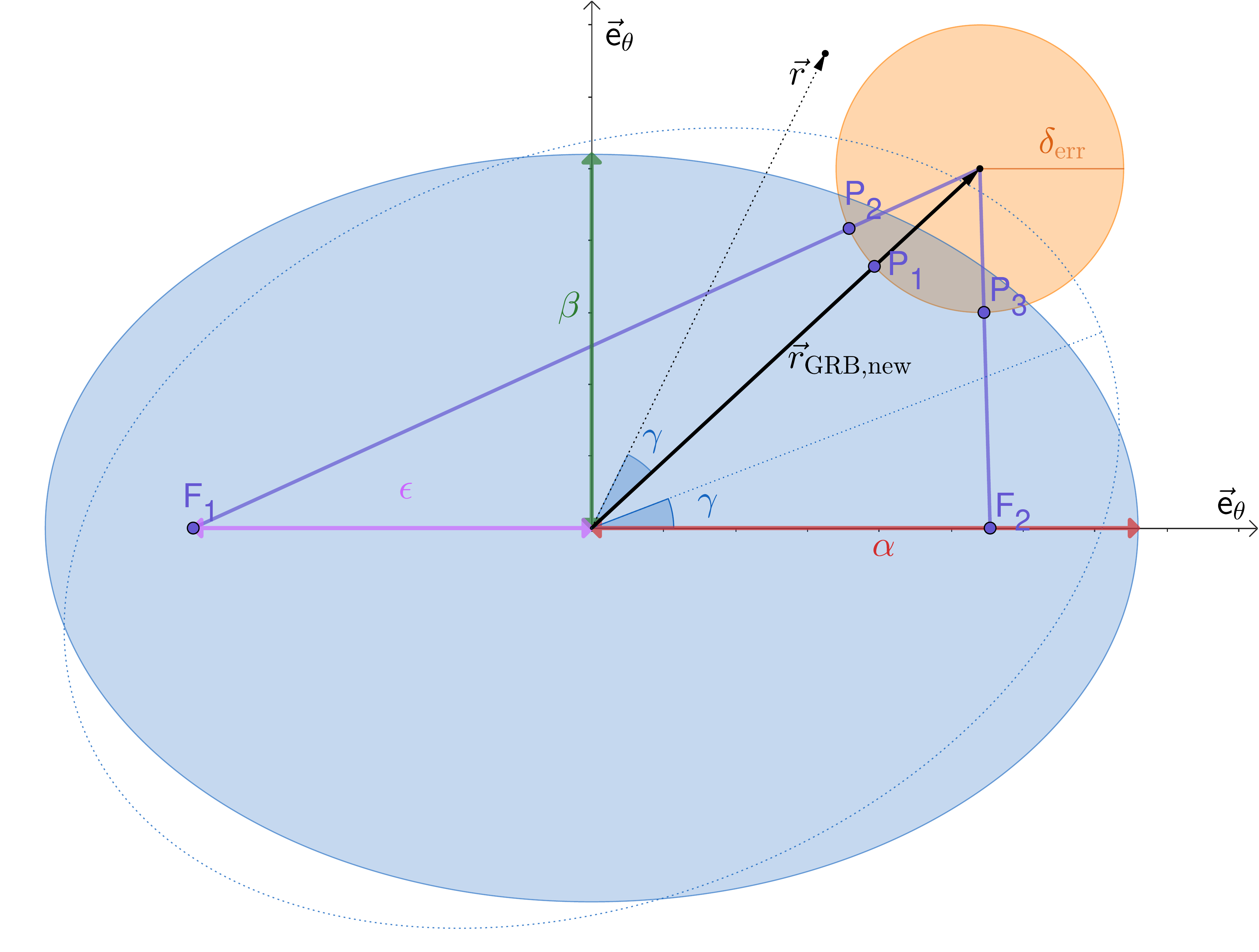}
  \includegraphics[width=0.49\textwidth]{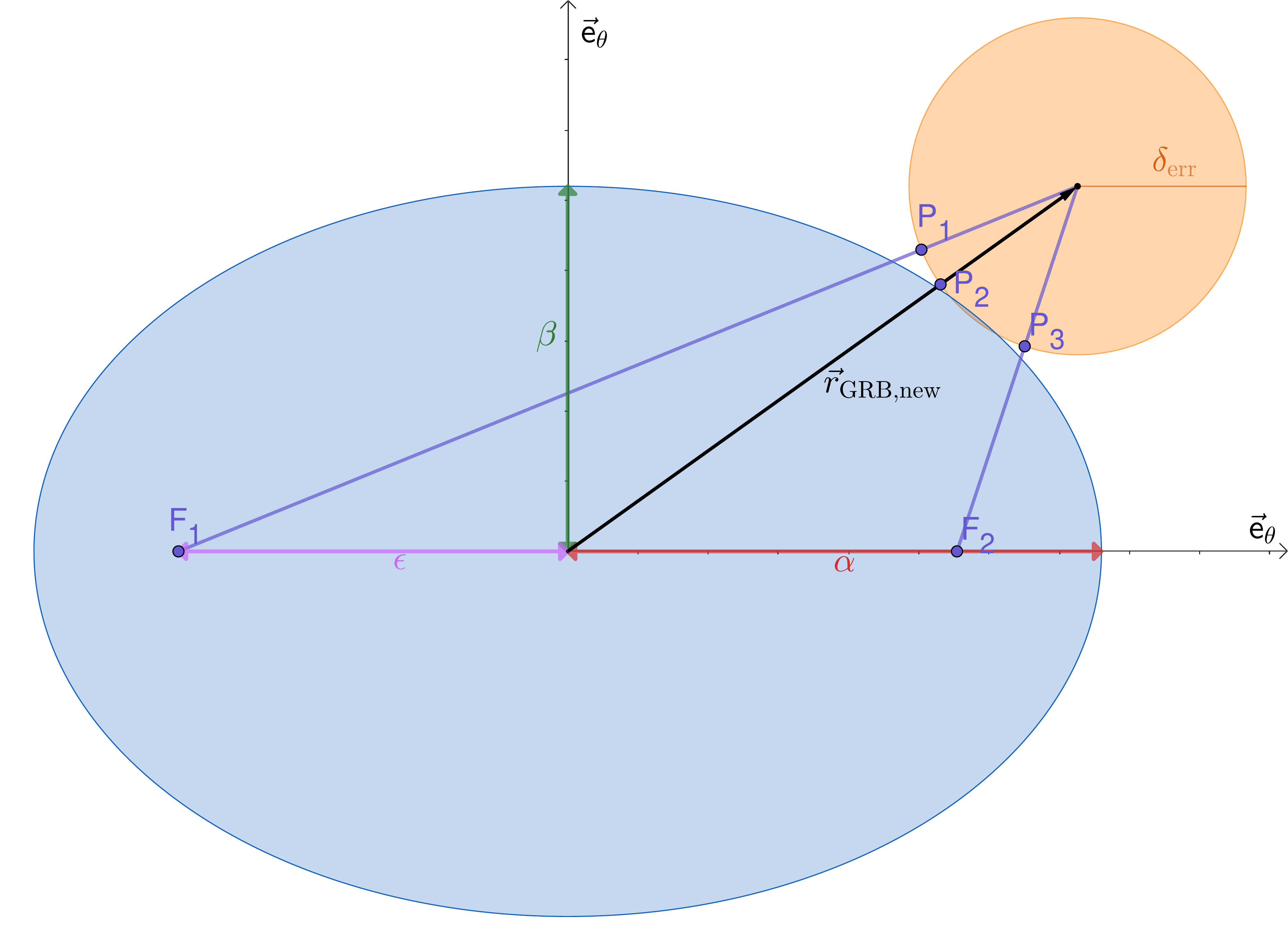}
	\caption{Two-dimensional illustration of an elliptical galaxy (blue) and uncertainty circle of a GRB (orange).
  The shape of the elliptical galaxy is given by the angular minor axis $\alpha$ and major axis $\beta$.
  $F_1$ and $F_2$ denote the focus points of the ellipse.
  The galaxy's orientation in the $\vec{e}_\phi$ and $\vec{e}_\theta$ plane is given by the angle $\gamma$.
  To account for the orientation of the galaxy on the sky, given by $\gamma$,
  the vector $\vec{r}$, that is highlighted with a dotted line, is rotated by
  the same angle in the opposite direction, yielding
  $\vec{r}_{\mathrm{GRB,new}}$. \textbf{Left:} Case in which the galaxy's
  ellipse and the GRB's error circle intersect, all three points $P_1$, $P_2$
  and $P_3$ lie within the galaxy's ellipse, and thus an association of the GRB
  with the galaxy is considered. \textbf{Right}: Exceptional case in which all
  three intersection points $P_1$, $P_2$, and $P_3$ lie outside the ellipse but
  the circle and ellipse have a small intersection area nonetheless. An
  association between the GRB and the galaxy is excluded.}

 \label{fig:ellipse_with_unc_2d}
\end{figure*}

When a GRB triggers a detector, its angular coordinates
$\phi_{\mathrm{GRB}}, \, \theta_{\mathrm{GRB}}$ are determined within
uncertainties. Without knowing the distance to the GRB, the host galaxy of the
GRB can never be securely determined. From the perspective of an observer on
Earth, the GRB coincides with the location of a galaxy if the projected GRB's
uncertainty region overlaps with the elliptic galaxy (see
Fig. \ref{fig:ellipse_with_unc_2d}).

Using spherical coordinates, the ellipse equation of a galaxy that is centered
at the origin, is symmetric around the $\vec{e}_r$ axis, and has its major and
minor axis aligned with the $\vec{e}_\phi$ and $\vec{e}_\theta$ axis, is given
by
\begin{equation}
  \left(\frac{\phi}{\alpha}\right)^2 + \left(\frac{\theta}{\beta}\right)^2 \leq 1.
  \label{eq:cone}
\end{equation}
To find out whether the GRB's uncertainty circle and galaxy's ellipse intersect,
we first evaluate the relative two-dimensional vector
\begin{equation}
  \vec{r} = \vec{r}_{\mathrm{GRB}} - \vec{r}_{\mathrm{cen}}
\end{equation}
between the galactic center $\vec{r}_{\mathrm{cen}}$ and the GRB's position
$\vec{r}_{\mathrm{GRB}}$.  The orientation of the galaxies on the sky is given
by their rotation angle $\gamma$ that is defined in the $\vec{e}_\phi$,
$\vec{e}_\theta$ plane (see Fig. \ref{fig:ellipse_with_unc_2d}). As the rotation
angle is not given in the catalog, we sample it for each galaxy from a uniform
distribution between $0^\circ$ and $180^\circ$ with a fixed seed and thus keep
the orientation constant in all simulated universes. As it is easier to rotate a
vector instead of changing the ellipse equation, the vector $\vec{r}$ is rotated
in the opposite direction by the angle $\gamma$, yielding the vector
$\vec{r}_{\mathrm{GRB,new}}$. If the GRB's rotated relative vector
$\vec{r}_{\mathrm{GRB,new}}$ fulfills Equation \eqref{eq:cone} for a galaxy, we
assume an association between the GRB and the galaxy and select the GRB for
further studies.

Assuming the GRB's location is measured within a circle of radius
$\delta_{\mathrm{err}}$, the GRB is associated with an elliptic galaxy if the
error circle of a GRB and the elliptical galaxy intersect. So, in addition to
the center of the GRB's location, we test if at least one of three more points,
$P_1$, $P_2$ or $P_3$, lie within the galaxy's ellipse (see
Fig. \ref{fig:ellipse_with_unc_2d}). We define the point $P_1$ as the
intersection point between the line connecting the galaxy's center with the
GRB's center (see Fig. \ref{fig:ellipse_with_unc_2d}). The points $P_2$ and
$P_3$ are defined by the intersection of the line between the focus points,
$F_1$ and $F_2$, and the center of the rotated GRB, given by
$\vec{r}_{\mathrm{GRB,new}}$, with the GRB's outer radius. The distance of the
two focus points to the center of the ellipse is given by
$\epsilon = \sqrt{\alpha^2-\beta^2}$.

Note here, that there are cases in which all three points lie outside the
ellipse but the ellipse and the circle intersect nonetheless (see right panel in
Fig. \ref{fig:ellipse_with_unc_2d}). As this is only the case when the
intersection area between the ellipse and circle is small, we neglect these edge
effects because an association is very unlikely to be made when the intersection
is so weak.

\subsection{GBM Trigger Selection}
To find out which of the sGRBs in the simulated universe would be detected by
GBM, the trigger algorithm of GBM is applied to the simulated GRB data.

A detailed overview of GBM's trigger algorithm is given in \citet{GBM2009} and
\citet{2012Paciesas}.  GBM's detectors are triggered when in at least two
detectors an increase of the measured background subtracted rate is measured
that is above a specified threshold.  The significance $s$ is defined as
\begin{equation}
  s = \frac{N_\mathrm{s} - \Delta t_\mathrm{s}\frac{N_\mathrm{b}}{\Delta t_\mathrm{b}} }{\sqrt{\Delta t_\mathrm{s}\frac{N_b}{\Delta t_\mathrm{b}} }}.
\end{equation}
Here, $N_\mathrm{s}$ is the number of counts measured in the source time
interval $\Delta t_\mathrm{s}$, and $N_\mathrm{b}$ is the number of background
photons that were detected in the previous time interval $\Delta t_\mathrm{b}$.
The background rate $N_\mathrm{b}/\Delta t_\mathrm{b}$ is defined as the average
rate that is obtained from the 17 s before the start of the tested source time
interval.  As threshold, $s>4.5$ is set as default in \texttt{cosmogrb}.  Note
here, that the chosen significance definition was shown to yield incorrect
results \citep{1983Li}. Nonetheless, we use it to imitate GBM's onboard
trigger algorithm.

The significance is computed for four different energy ranges, (1) $50-300$ keV,
(2) $25-50$ keV, (3) $>100$ keV, and (4) $>300$ keV.  For the energy ranges (1)
and (2), ten time scales for $\Delta t_\mathrm{s}$ between 16 ms and 8.192 s
were tested, successively increasing as $16 \, \mathrm{ms} \cdot 2^{n}$ with
$n\in [0,1,...,9]$.  The tested time scales for (3) include only the first nine
time intervals of (1) and (2).  For the energy range defined by (4), the
significance is only computed for the first four time intervals.

\subsection{Star Formation Rate Selection}
As magnetars are thought to originate from core collapse supernovae, and
considering that MGFs can only be produced by young magnetars, it is expected
that MGFs originate preferably in galaxies with a large star formation rate
(SFR), similar to long GRBs. Approximately 64 \% of the galaxies in the LV
catalog have a measured value for the SFR, either originating from the
H$_\alpha$ line emission or the far ultraviolet (FUV) continuum luminosity.

The galaxy with the smallest SFR that was proposed in previous work as host
galaxy for MGFs is the Andromeda Galaxy with an H$_\alpha$ SFR of
$0.5 \, M_\odot \, \mathrm{yr}^{-1}$ and FUV SFR of
$1.0 \, M_\odot \, \mathrm{yr}^{-1}$.  These two values were used as thresholds
for the galaxies.  All coinciding galaxies with an SFR that is smaller than the
given values by the Andromeda Galaxy are discarded, as they are assumed unlikely
to host MGFs.  If no value for the SFR is given in the LV galaxy catalog, or an
upper limit that is larger than the threshold, the galaxy is not excluded as
potential host association.

\subsection{Duration Selection}
MGFs are expected to be short pulses, and are mainly searched for in catalogs of
sGRBs which are traditionally defined with a duration less than 2 s.  As we
sample the duration of the simulated sGRBs from the full, non-truncated best-fit
distribution, the duration can also be larger than 2 s.  To mimic the selection,
another cut on the simulated GRBs can be performed by choosing only GRBs with
smaller durations than 2 s.

\subsection{Isotropic Energy Release Selection}
Another selection can be applied after computing the isotropic energy, which is
given by Eq. \eqref{eq:Eiso}.  When using the spatial selection, the GRB's host
is associated to be the found LV galaxy for the selected GRBs. Thus, for the
luminosity distance $d_L$ of the GRB, the distance to the corresponding
associated galaxy is used instead of the true distance, given by the sampled
redshift. Assuming that there is a maximum on the magnetic field strength in a
magnetar that can power an MGF, the measured maximum isotropic energy release,
$E_{\mathrm{iso, max}}=5.3 \times 10^{46}$ erg of ${\mathrm{MGF} \, 051103}$ \citep{Ofek2006},
is used as a threshold for the selection of MGF candidates.

\subsection{Fluence Selection}
If a GRB has a localization error in the order of degrees, an association with a
specific galaxy becomes unlikely. All previously proposed extragalactic MGFs
were detected with multiple instruments in the Interplanetary Network (IPN).With
the triangulation method, the localization of the triggered events can be
improved to yield error boxes in the arcmin or even arcsec range. The typical
error for the measurements of $\gamma$-rays with GBM alone are in the order of
degrees. The detection of GRBs in the X-rays with Swift/BAT can also yield
localizations in the arcmin range but the instrument detects on average 2.5
times less sGRBs than Fermi/GBM, which is why we do not focus on
Swift observations in this work. A detailed summary of the localization methods
of all current and future $\gamma$-ray instruments, and their accuracy, is given
in \citet{2022Greiner}.

So, when simulating GBM sGRBs with a small or no localization error, this would
be only feasible if an IPN localization exists. However, not for all sGRBs that
are detected by GBM, a localization with the IPN is possible. The detection
probability of all instruments in the IPN scales with the fluence of the source,
which denotes the time-integrated measured flux in a specific energy
interval. The IPN efficiency is equal or larger than 50 \% for a fluence equal
or larger than $10^{-6}$ erg cm$^{-2}$ \citep{2013Hurley}.  The efficiency was
computed in \citet{2013Hurley} for two cases, in the first for the measurement
of the $\gamma$-rays with any two spacecrafts of the IPN, and in the other for
two widely spaced instruments. The precision of the localization with
triangulation is proportional to the distance of the spacecrafts. We only
consider the case of two widely separated spacecrafts as only in this case can
the localization error be efficiently decreased.  We make a final selection by
using the fluence threshold of >$10^{-6}$ erg cm$^{-2}$ to choose the sGRBs with
a high potential for a localization with the IPN and a decreased localization
error.

\section{Model checking}\label{sec:model_checks}

We performed three tests to verify that the population synthesis produces
results that are in agreement with observations of sGRBs. To do so, we simulate
a full survey of GRBs corresponding to an observation time frame of 14 years
(the active time of GBM) in a comoving volume of up to redshift $z=5$.

In the first test (Sect. \ref{sec:number}), our aim is to determine whether the
number of simulated sGRBs which are selected by the simulated GBM trigger is in
agreement with the number of detected GRBs during the observation time of 14
years with GBM. Afterwards (Sect. \ref{sec:pop_params}), we show the sGRB
population parameter distributions to validate that the sampled parameters
follow the chosen distributions. In the third test
(Sect. \ref{sec:recover_pop_params}), we aim to check whether the input spectral
parameters for each GRB can be recovered when using the previously described
analysis methodology.

\subsection{Number of GRBs}\label{sec:number}

Integrating the spatial distribution as in Equation \eqref{eq:volumeintegral}
and using the parameter values of case (c) as in \citet{Ghirlanda2016} with
$\dot{\rho_0} \Delta t_{\mathrm{obs}}=11.2 \, \mathrm{Gpc}^{-3}$ yields the
volume integral of 1676.3. As the measurement of GRBs in a given time interval
is equivalent to a counting experiment, the number of GRBs in a survey is drawn
from a Poisson distribution with the expected value given by the volume integral
result $\lambda = 1676.3$.

In the test universe, 1623 GRBs were simulated of which only 157 were selected
by the implemented GBM trigger. In contrast, the median best-fit lognormal
distribution of the GRBs' duration $T_{90}$ suggests that there are around 724
sGRBs in the GBM sample. Thus, almost five times more GRBs were detected in the
past 14 years than our simulations predict. The reason for the mismatch between
the number of detected GRBs from the simulated sample in comparison to the
observed GBM sample is most likely due to an underestimated normalization
constant of the spatial distribution used in \citet{Ghirlanda2016}. Previous
predictions for the redshift distribution of sGRBs differ by a factor of 300
(comparing results of \citet{2013Dominik} and \citet{Ghirlanda2016}). Even the
two different assumptions of the $E_\mathrm{p}$-$E_{\mathrm{iso}}$ and
$E_\mathrm{p}$-$L_{\mathrm{iso}}$ correlation, and the independent probability
distribution in \citet{Ghirlanda2016}, yield values for the normalization
constant that differ by a factor of four.

The reason for the underestimated value of the normalization constant in
\citet{Ghirlanda2016} is most likely based on selection effects. Instead of
using a forward-folding model as we are using, \citet{Ghirlanda2016} assume that
their chosen sample of sGRBs is representative for the entire population of
sGRBs.

Therefore, we increased the normalization constant in the following analysis to
$\dot{\rho_0} \Delta t_{\mathrm{obs}}=57\, \mathrm{Gpc}^{-3}$ so that the number
of detected sGRBs is in a similar range as the measured number of sGRBs (724) in
the lifetime of GBM.

\subsection{Population Parameter Distributions}\label{sec:pop_params}
\begin{figure*}[htpb]
  \centering
  \begin{tikzpicture}
    \node[anchor=south west,inner sep=0] (image) at (0,0)
    {\includegraphics[width=0.99\textwidth]{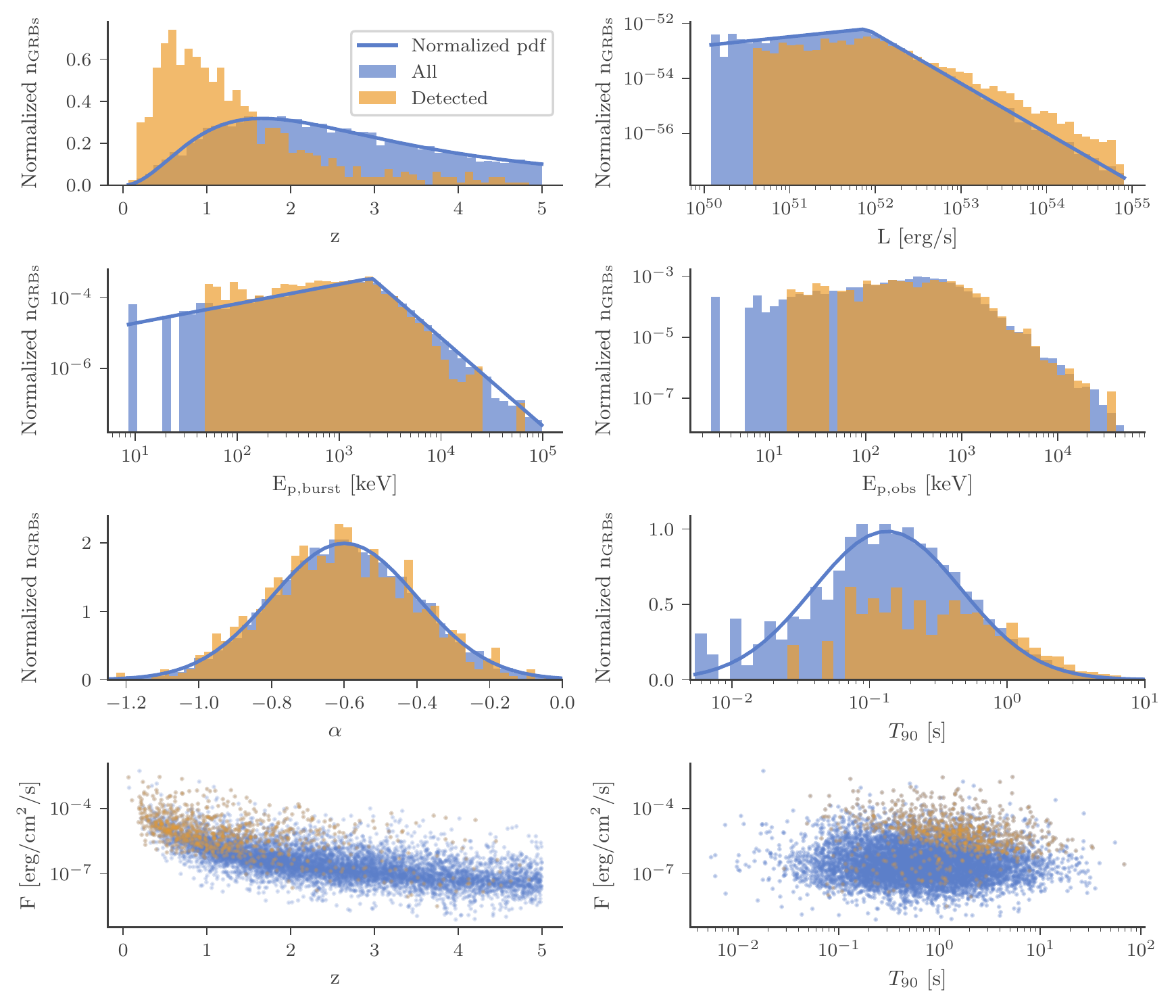}};
    \begin{scope}[ x={(image.south east)}, y={(image.north west)} ]
      \node [black, font=\bfseries] at (0.04,0.999) {(a)}; \node [black,
      font=\bfseries] at (0.51,0.999) {(b)}; \node [black, font=\bfseries] at
      (0.04,0.75) {(c)}; \node [black, font=\bfseries] at (0.51,0.75) {(d)};
      \node [black, font=\bfseries] at (0.04,0.505) {(e)}; \node [black,
      font=\bfseries] at (0.51,0.505) {(f)}; \node [black, font=\bfseries] at
      (0.04,0.25) {(g)}; \node [black, font=\bfseries] at (0.51,0.25) {(h)};
    \end{scope}
  \end{tikzpicture}
   \caption{Normalized density histograms of the simulated redshift $z$ \textbf{(a)}, luminosity $L$ \textbf{(b)}, peak energy $E_\mathrm{p,burst}$ in the burst frame \textbf{(c)} and $E_\mathrm{p,obs}$ in the observer frame \textbf{(d)}, the power-law index $\alpha$ \textbf{(e)} and the duration $T_{90}$ \textbf{(f)}.
   Solid lines display the chosen probability density functions as described in Section \ref{sec:popsynth}, using the parameters corresponding to case (c) of \citet{Ghirlanda2016}, and an increased value for the redshift normalization constant of $\dot{\rho_0} \Delta t_{\mathrm{obs}}= 57 \, \mathrm{Gpc}^{-3}$
   \textbf{Blue}: All simulated GRBs. \textbf{Orange}: From GBM trigger selected
   GRBs. The histograms are normalized by dividing each bin by the bin width in
   such a way that their integral yields unity. The lower panels \textbf{(g)}
   and \textbf{(h)} show the scatter of the simulated flux as a function of the
   redshift and duration on the left and right respectively.}
 \label{fig:distr_sgrbs}
\end{figure*}
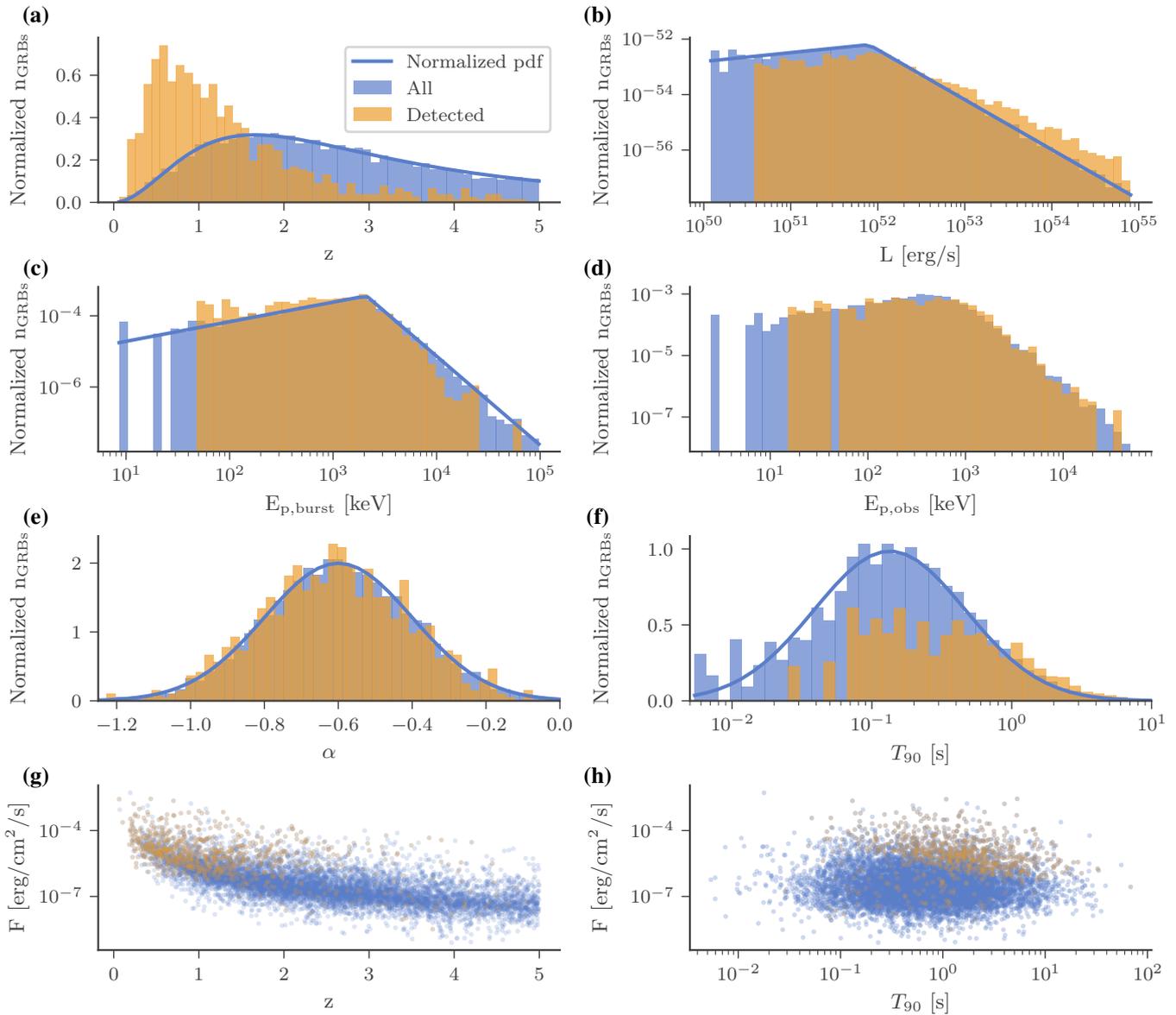

Using the increased normalization constant of the redshift distribution, the
normalized histograms for all parameters characterizing a full universe of sGRBs
are shown in Fig. \ref{fig:distr_sgrbs}. In this test universe, 8440 GRBs were
simulated, and 805 were selected by the GBM trigger. This is in a similar range
as the number of detected sGRBs with GBM.
As visible in Fig. \ref{fig:distr_sgrbs}, the sampled parameters follow the
chosen probability density functions. As expected, there is a bias towards the
observation of sGRBs with a small distance, namely small redshift (see panels
(a) and (g) in Fig. \ref{fig:distr_sgrbs} ) which is due to the measured flux
scaling by the luminosity distance squared which is dependent on redshift.  The
farther away an sGRB is, the dimmer it is and the more unlikely is its
detection. We find another bias on the detection of bright sources with larger
luminosity and flux (see Fig. \ref{fig:distr_sgrbs} (b) and (g)), which agrees
with expectations too. The brighter a source is, the more likely its detection
is. Furthermore, GRBs with a preferably longer duration are detected (see
Fig. \ref{fig:distr_sgrbs} (f)) which is due to GBM's trigger algorithm, and its
limited time resolution.

\subsection{Recovering Population Parameters from Simulated GBM Data}\label{sec:recover_pop_params}
In addition, we tested whether the input latent parameters of the sGRBs can be
recovered when analyzing the simulated GBM data as described in Section
\ref{sec:threeml}, and thus if our simulations are consistent. Not for all GRBs
that trigger GBM can a spectrum be fit. The more source counts are measured
above the background, the more likely it is that the fit converges. It was found
that for approximately 75 \% of the simulated GRBs that were identified by the
GBM trigger algorithm as detected, a spectrum could be fit. For about 90 \% of
the sGRBs for which the fitting process failed, the Bayesian blocks technique
could not determine a source time interval. For the other sGRBs, the fitting
algorithm did not converge and thus failed.

We randomly choose a detected GRB from the test universe which is representative
for the other simulated and detected sGRBs for which a spectrum was fitted
successfully. The chosen simulated GRB is at redshift $z=0.60$ and has a
duration of 0.65 s. Its sampled flux is $8.12 \times 10^{-6}$ erg cm$^{-2}$
s$^{-1}$. The latent spectral parameters of the chosen GRB's spectrum are the
peak energy in the burst frame $E_{\mathrm{p}} = 788$ keV and the power law
index $\alpha=-0.82$. In the observer frame, the peak energy is shifted to lower
energies $E_{\mathrm{p,obs}}=E_{\mathrm{p}}/(1+z)$, and is thus at
$E_{\mathrm{p,obs}}=486.25$ keV.

Using \texttt{gbmgeometry}, the closest BGO and the three closest NaI detectors
can be determined which have the smallest separation angle between their
pointing and the direction of the GRB. For the test GRB, the closest BGO
detector is $b_0$, and the closest NaI detectors are in descending order $n_1$,
$n_0$ and $n_3$.

\begin{figure}[htpb]
  \centering
  \includegraphics[width=0.49\textwidth]{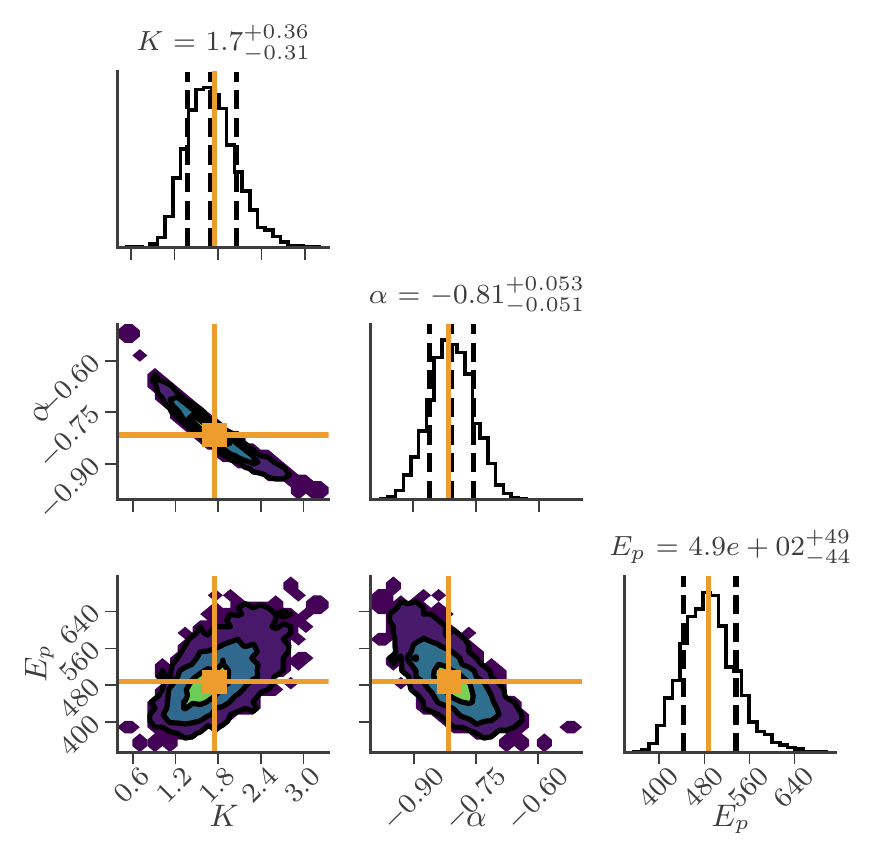}
  \caption{Corner plot of the CPL fit of the test GRB. The CPL is parameterized
    as in Equation \eqref{eq:cpl} with $E_{\mathrm{piv}}= 1 $ keV. The peak
    energy $E_{\mathrm{p}}$ has units of keV, the power law index is unit-less
    and the normalization $K$ has units of cm$^{-2} $ s$^{-1}$ keV$^{-1}$. The
    true chosen parameters are highlighted in orange.}
  \label{fig:grb1_bestfit_cornerplot}
\end{figure}

With the Bayesian blocks technique, the source interval was found to have a
duration of 0.65 s, which is the same as the latent duration of the sGRB. A CPL
spectrum as parameterized in Equation \eqref{eq:cpl} is fit with \texttt{3ML} to
the counts that were measured in the set source interval. The corner plot of the
CPL fit for the chosen GRB is shown in
Fig. \ref{fig:grb1_bestfit_cornerplot}. The modes of the marginal posterior
distributions of the fit match the latent parameters.

\section{Simulations and Results}\label{sec:results}
First, we studied how the number of spatial coincidences changes as a function
of the angular error radius of the sGRB localization. Then, we assume that the
localization of the sGRBs has no uncertainties, corresponding to the existence
of an IPN localization with errors in the order of arcmin or even arcsec. We
simulate 1000 surveys of sGRBs with a constant temporal profile. For the
redshift and luminosity distribution, we use the parameters of \citet
{Ghirlanda2016}, as summarized in Table \ref{tab:ghirlanda_fit_params}, except
that we use the increased normalization constant
$\dot{\rho_0} \Delta t_{\mathrm{obs}}=57\, \mathrm{Gpc}^{-3}$ for the redshift
distribution, as described in Sec. \ref{sec:number}. For the GRBs coinciding
with a LV galaxy, we simulate GBM data, run the GBM trigger algorithm, and
search for MGF candidates, based on the selection criteria that we described in
Sec. \ref{sec:selection_criteria}. The results of these simulations are
summarized in the following.

\subsection{Spatial Selection}\label{sec:results_nolocerror}
\begin{figure*}[htbp]
  \centering
  \includegraphics[width=0.99\textwidth]{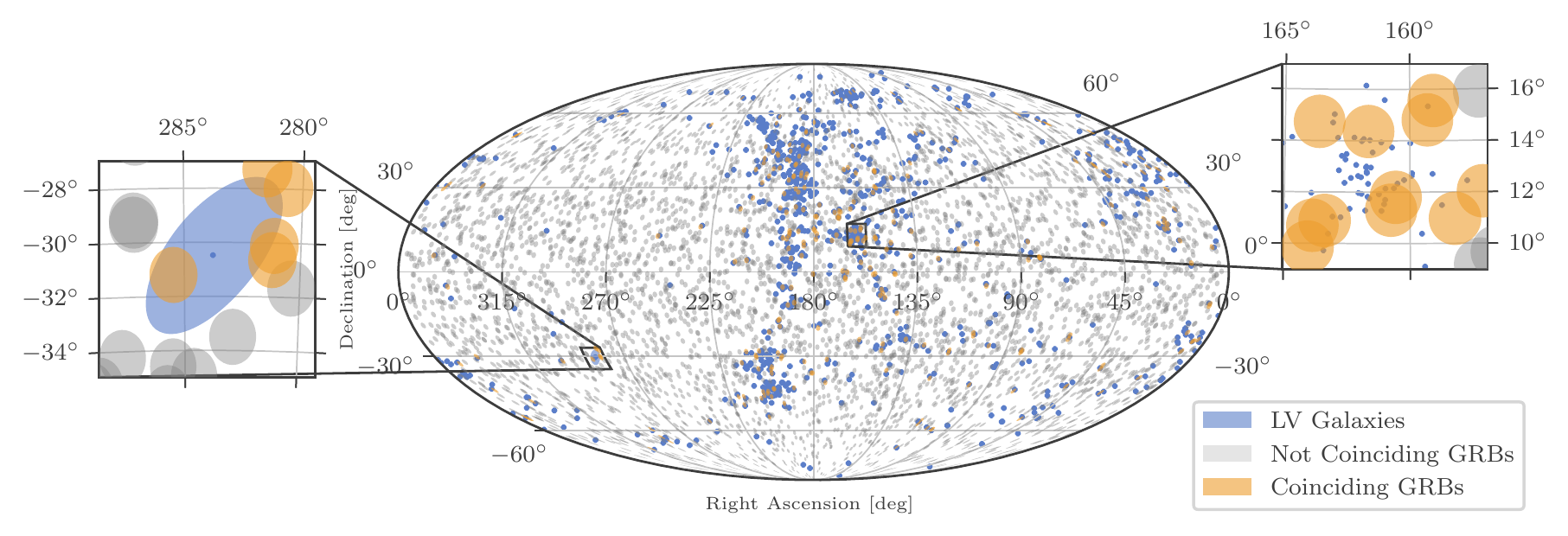}
  \caption{Visualization of the locations of sGRBs on the sky in a simulated
    survey, in which an angular error radius of 1$^\circ$ for the localization
    of all sGRB is assumed. The coordinates are given in the International
    Celestial Reference System (ICRS). sGRBs coinciding with the position of a
    Local Volume Galaxy are highlighted in
    orange.}
  \label{fig:survey_spatial_selec}
\end{figure*}

To study how the number of spatial coincidences changes as a function of the
error on the sGRB localization, we simulate 50 surveys of sGRBs assuming
different uncertainty radii around the simulated true location of the sGRBs.  We
compute the number of spatial coincidences in each of the surveys.  In the left
panel of Fig.\ref{fig:survey_spatial_selec}, one simulation of a population
of sGRBs with an error circle of radius 1$^\circ$ is shown. In the example, 8440
sGRBs were simulated, of which 337 could be associated with at least one galaxy
of the LV catalog.

\begin{figure}[htbp]
  \centering
  \includegraphics[width=0.49\textwidth]{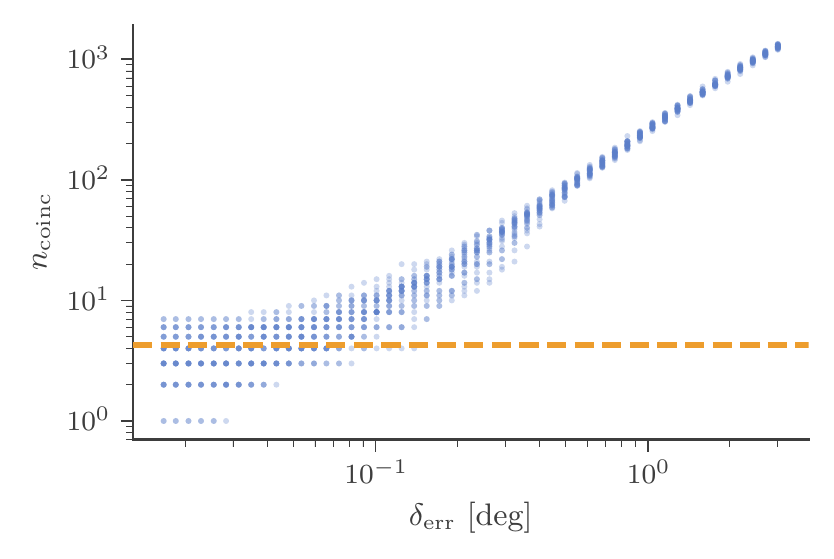}
  \caption{Number of coincidences of the GRB's location with a galaxy in
    simulated surveys as function of the GRB's error circle radius
    $\delta_\mathrm{err}$.  The scatter originates from the results of 50
    surveys of sGRBs that were simulated for each chosen error circle. The
    orange line corresponds to the expected number of spatial coincidences at
    small error radii (see Eq. \eqref{eq:exp_n_coinc})}.
  \label{fig:frac_coinc_err}
\end{figure}

The number of spatial coincidences in the simulated surveys as a function of the
uncertainty of the GRB localization is shown in Figure
\ref{fig:frac_coinc_err}. Up to an error radius of about
{0.03$^\circ=1.8 \, \mathrm{arcmin}$}, the uncertainty of the GRB's localization
does not affect the number of coincidences significantly and is thus independent
of the localization error. As the positions of the GRBs are sampled in such a
way that they are isotropically distributed at the sky, the number of spatial
coincidences is expected to be only dependent on the angular fraction of the sky
that is covered by the nearby galaxies and the number of GRBs in the
survey. Thus, the expected number of coincidences is given by
\begin{equation}
  n_{\mathrm{coinc, \, exp}} = N \, \frac{\sum_{i=1}^{N_{\mathrm{LV}}} \pi \, a_i \, b_i }{4\pi},
  \label{eq:exp_n_coinc}
\end{equation}
where $N$ is the expected number of sGRBs in a survey which is given by the
volume integral in Eq. \eqref{eq:volumeintegral}. As visible in Fig.
\ref{fig:frac_coinc_err}, the number of spatial coincidences scatters around the
expected number $n_{\mathrm{coinc, \, exp}}$.

For larger error radii, the number of spatial coincidences rises as a power of
the circular error radius. By increasing the error radius by one order of
magnitude from $0.1^\circ$ to 1$^\circ$, the number of coincidences rises
approximately by a factor of 40, so more than an order of magnitude.

If the localization error is large, an association of a GRB with a specific
galaxy becomes unlikely as there are too many galaxies in the field of view.
The localization region of all previously propsed MGF candidates was in the
order of arcmin as they all had an IPN localization.  This corresponds to the
region in which the number of spatial coincidences is independent of the GRB's
localization error. In the following, we simulate 1000 surveys in which we set
the error radius to zero, thus assuming that there is an IPN localization for
all sGRBs.  We select only the sGRBs that are spatially coinciding with one of
the LV galaxies.

\subsection{SFR Selection}
We apply a selection on the coinciding galaxies based on their SFR. All
coinciding galaxies with an SFR that is smaller than the SFR by the Andromeda
Galaxy are discarded, as they are assumed unlikely to host MGFs. If no value for
the SFR is given in the LV galaxy catalog, or an upper limit that is larger than
the threshold, the galaxy is not excluded as potential host association.

Of all 1244 galaxies in the LV catalog, 162 different galaxies were at least
once aligned with one of the sGRBs in the 1000 simulated surveys. Only 44 of the
162 galaxies (about 27 \%) fulfill the SFR selection criterion, and can thus be
considered as potential host galaxies for MGFs.

\begin{figure}[htb]
  \centering
  \includegraphics[width=0.49\textwidth]{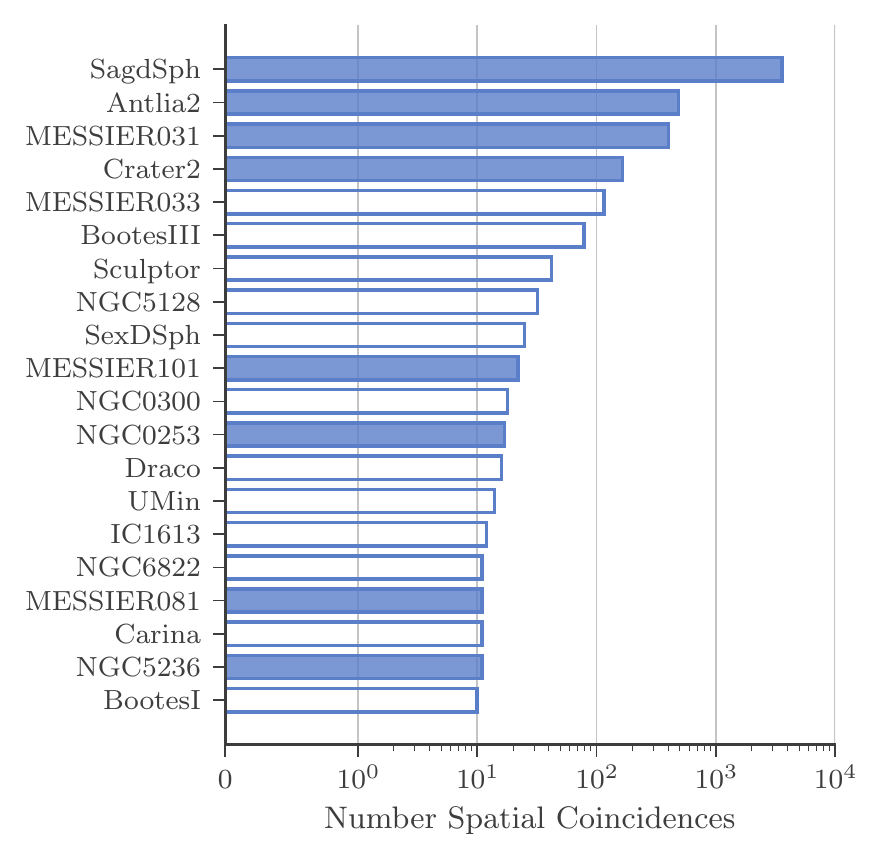}
  \caption{Number of spatial coincidences summed over all surveys of GRBs for
    the 20 most often coinciding galaxies in all 1000 simulated surveys.
    Galaxies fulfilling the SFR selection criteria are highlighted by filled 
    bars. Excluded galaxies are illustrated by unfilled bars.}
  \label{fig:coincidences_gals_sfr1}
\end{figure}

In Fig.\ref{fig:coincidences_gals_sfr1}, a histogram of the 20 galaxies that
most often align with the simulated sGRBs are shown. The number of spatial
coincidences is summed over all simulated surveys. The galaxies fulfilling the
SFR criterion are highlighted by filled bars. While in the simulations, there
was no chance coincidence of sGRBs with the galaxies M82 and M83, the other two
galaxies (M31, NGC0253) that were proposed as hosts for the observed MGFs 070201
and 200415A belong to the 12 most often coinciding galaxies. Note here that in
the literature, the galaxy associated with {200415A} is called Sculptor but is
named NGC0253 in the LV catalog. The galaxy named Sculptor in the LV catalog is
a different galaxy which is excluded as host for MGFs because of its small SFR.

\begin{figure}[htb]
  \centering
  \includegraphics[width=0.49\textwidth]{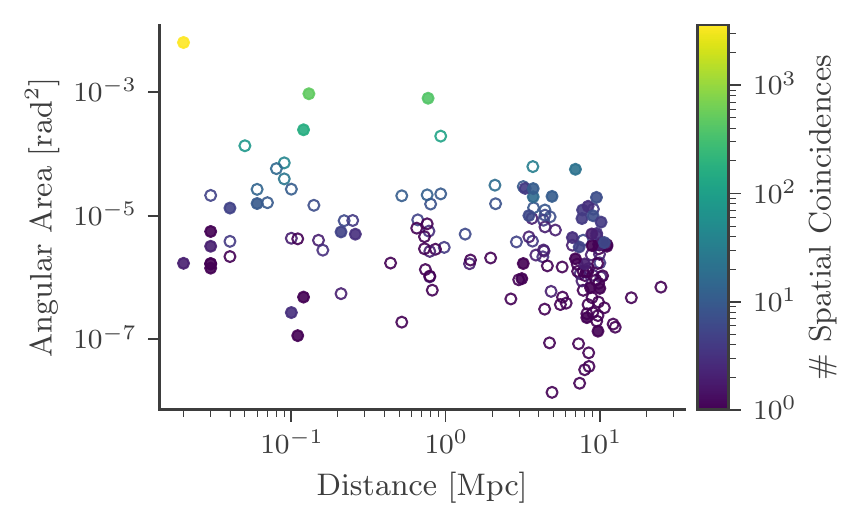}
  \caption{Number of spatial coincidences summed over all surveys as a function
    of the angular area and distance of the galaxies. Galaxies fulfilling the
    SFR selection criteria are highlighted by filled circles. Excluded
    galaxies are illustrated by unfilled circles.}
  \label{fig:coincidences_gals_sfr2}
\end{figure}

How often an sGRB coincides with a galaxy is proportional to its angular size.
The number of spatial coincidences as a function of the galaxies' angular area
and distance is shown in the right panel of Figure
\ref{fig:coincidences_gals_sfr2}. By far the most spatial coincidences are in
association with the Sagittarius Dwarf Galaxy (SagdSph). In comparison to the
second most often associated galaxy, Antlia 2, the number of associations with
the Sagittarius Galaxy is larger by an order of magnitude, and thus dominates
the result. The Sagittarius Galaxy is also by far the largest galaxy on the sky
after excluding the Small and Large Magellanic Cloud from the LV catalog. As
there is no constraint on the SFR given for the Sagittarius Galaxy, we do not
remove the galaxy from the LV sample.

\subsection{GBM Trigger Selection}\label{sec:gbm_trigger_selec}
\begin{figure}[htb]
  \centering \includegraphics[width=0.49\textwidth]{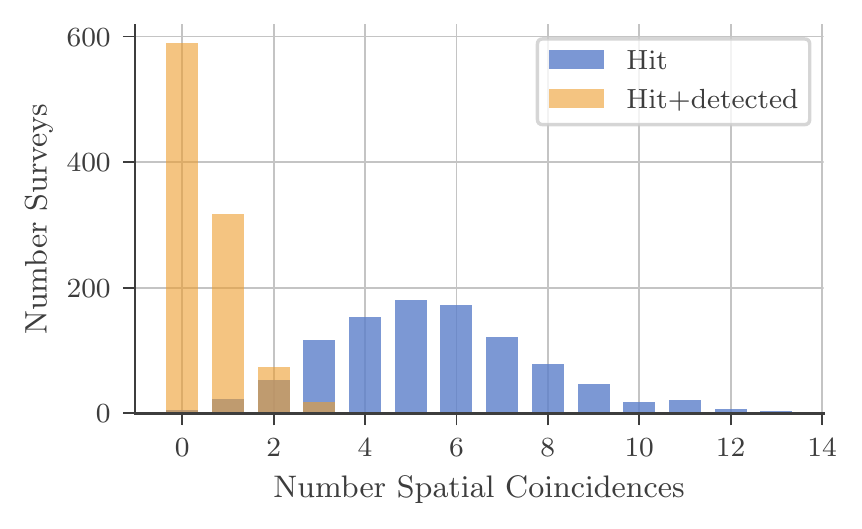}
  \caption{\textbf{Blue}: Number of surveys in which a specific number of sGRBs
    coincided with the location of a nearby galaxy. \textbf{Orange}: sGRBs that
    were selected by the GBM trigger algorithm in addition to coinciding with a
    galaxy.}
  \label{fig:coincidences}
\end{figure}

In Fig.\ref{fig:coincidences}, it is shown in how many surveys we find a
specific number of spatial coincidences of sGRBs and LV galaxies. In 99.9 \% of
all simulated sGRB surveys, at least one of the GRBs coincides with one of the
galaxies in the LV catalog. Of the 5469 sGRBs in all 1000 simulated surveys that
coincide with a LV galaxy, 524 are detected by the GBM trigger algorithm.  There
are 411 surveys in which at least one detected GRB coincides with a nearby
galaxy. Thus, when only considering the spatial and GBM trigger selection, there
is a chance of about 41 \% that in the observing duration of 14 years, there is
at least one sGRB detected that can be falsely associated with a nearby galaxy.

The number of coincidences is directly proportional to the observing time as the
number of GRBs in the universe is given by the integrated redshift distribution,
which is proportional to the observation time (see Equation
\eqref{eq:volumeintegral}). Thus, it can be computed after which observation
time the probability is larger than 5 \% to observe an MGF.
After a detection time of approximately 1.7 years, GBM is triggered by an sGRB
which is aligning by chance with a nearby galaxy with a probability of 5\%. Of
the 45 different galaxies that are found to be coinciding with detected sGRBs in
all simulated surveys, 17 fulfill the SFR condition.  The result is strongly
dominated by the Sagittarius Galaxy.

When combining the spatial, GBM trigger, and SFR selection, we find coincidences
in 298 surveys. Therefore, after 2.3 years, GBM observes with a 5\% chance an
sGRB that is coinciding with a nearby galaxy which has an SFR that is larger
than the one from the Andromeda Galaxy. The result is strongly dominated by the
Sagittarius Galaxy. Without the Sagittarius galaxy, GBM sees an sGRB with the
described properties after 6.8 years with a 5\% probability, and the detection
probability after 14 years is 10.3 \%.

\subsection{Duration Selection}\label{sec:dur_selec}
\begin{figure}[htpb]
  \centering \includegraphics[width=0.49\textwidth]{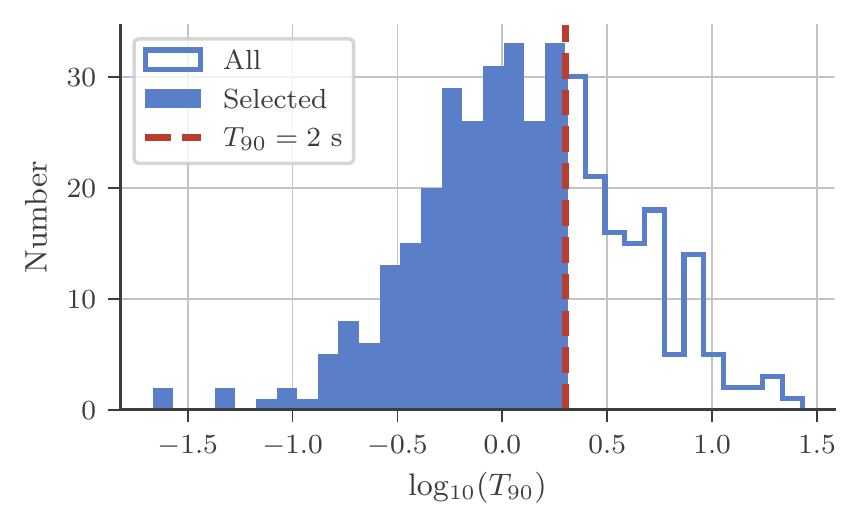}
  \caption{Histogram of the logarithmic durations of all sGRBs triggering GBM
    and coinciding with a LV galaxy. sGRBs with $T_{90}<2$ s are selected
    (filled histogram), the others are discarded (unfilled histogram). The
    threshold is highlighted by a red dashed vertical line.}
  \label{fig:selection_t90}
\end{figure}

The distribution of the durations of sGRBs triggering GBM and coinciding with
one of the LV galaxies is shown in Fig.\ref{fig:selection_t90}. The duration
was determined by using the Bayesian block method. For about 74 \% (386 sGRBs)
of the detected sGRBs, the Bayesian block method was able to determine at least
three time intervals, thus identifying a source interval in between the
background intervals. The other 26 \% of the sGRBs were discarded from the
following analysis. As there is a bias towards measuring sGRBs with longer
durations (see Fig.\ref{fig:distr_sgrbs}), there is a non-negligible number
of 148 sGRBs with a duration larger than 2 s.

After excluding the sGRBs with $T_{90}>2$ s, we still find at least one MGF
candidate in 226 surveys (22.6 \% of all simulated surveys).
When excluding the coincidences with the dominating Sagittarius Galaxy, in 85
surveys, there is at least one sGRB left that is selected by the spatial, GBM
trigger, and duration selection criteria.
A MGF candidate is found by chance after 3.1 years with 5 \% probability when
only considering events that trigger GBM and have a duration smaller than 2 s.

When combining the duration selection with the SFR selection, the detection
probability of an MGF candidate is 20.5 \%, without the Sagittarius Galaxy 6.2
\%. Hence, GBM is triggered with a non-negligible probability of 5 \% after
approximately 3.4 years by an sGRB that can be wrongly associated as MGF, and
when neglecting the Sagittarius Galaxy as host, after 11.3 years.

\subsection{Isotropic Energy Release Selection}\label{sec:Eiso_selec}
One of the main arguments for the candidacy of proposed extragalactic MGFs was
that the determined peak energy and isotropic energy release of the candidates
was not consistent with the measured population of sGRBs.  However, this relies
on the presumption that the distance to the events is that of the assumed nearby
host galaxy.  This results in the peak and isotropic energy being scaled to this
nearby distance rather than that of a possibly more distant origin.

\begin{figure}[htpb]
  \centering
  \includegraphics[width=0.49\textwidth]{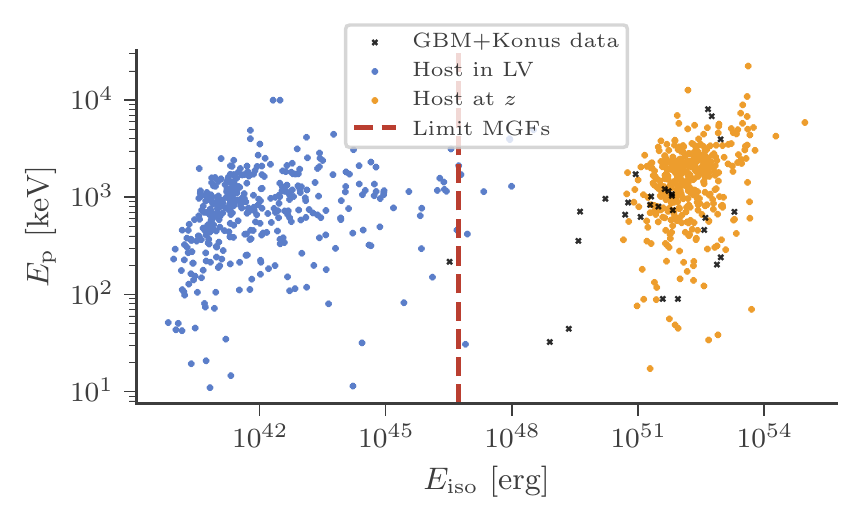}
  \caption{Scatter plot of the simulated peak energy $E_\mathrm{p}$ and
    isotropic energy release $E_\mathrm{iso}$ assuming a distance given by the
    coinciding LV galaxy (\textbf{blue}), and by the simulated redshift
    (\textbf{orange}). The black crosses correspond to the determined values for
    $E_\mathrm{p}$ and $E_\mathrm{iso}$ of sGRBs that were observed by GBM
    \citep{2021Poolakkil} and Konus Wind \citep{2017Tsvetkova}, respectively.}
  \label{fig:selection_ep_eiso_scatter}
\end{figure}

For all 386 simulated sGRBs which coincide with the location of a nearby galaxy,
which are selected by the GBM trigger algorithm, and for which the spectral fit
succeeded, we computed the isotropic energy release and peak energy. First, we
assume that the host of the GRB is at the distance given by the coinciding
nearby galaxy. Second, we assume that the latent redshift of all GRBs could be
correctly determined and scale the isotropic energy and peak energy to the
luminosity distance given by the redshift. The $E_\mathrm{p} $-$E_\mathrm{iso}$
scatter plot for both cases is shown in Figure
\ref{fig:selection_ep_eiso_scatter}. The results when using the latent redshift
are highlighted in orange, and for the assumed distance to the LV galaxies in
blue. In addition to the results of the simulated data, the results for
$E_\mathrm{p}$ and $E_\mathrm{iso}$ of all identified sGRBs by Konus Wind
\citep{2017Tsvetkova}, and GBM \citep{2021Poolakkil} are depicted. If the
redshift of the sGRBs is able to be correctly determined, $E_\mathrm{p}$ and
$E_\mathrm{iso}$ are in agreement with the measured sGRB population from Konus
Wind and GBM, as there are no far outliers. In contrast, if the distance to the
coinciding nearby galaxy is used, the points indeed look like a separate
population that is not in agreement with the sGRB population. Nevertheless, the
difference is only due to the extremely underestimated value for the distance of
the sGRBs, and is thus an artifact of the wrong host association.

Using the maximum value for the isotropic energy of ${5.3\times 10^{46} \, \mathrm{erg}}$ (the
highest ever derived for an MGF), 381 sGRBs are compatible with MGFs when setting
the distance to the ones of the coinciding galaxies. If in contrast the redshift
of the sGRBs is known, all sGRBs have an isotropic energy that is larger than
the limit given by the magnetic field strength, which is not compatible with the
interpretation as MGFs.

When using the spatial, GBM trigger, and isotropic energy selections for
identifying MGFs, we find in 30.4 \% of all simulated surveys, one or more
accidental MGF candidates. When excluding the Sagittarius Galaxy from the
analysis, we find accidental MGF candidates in 11 \% of the surveys. After 2.3
years of observing with GBM, an accidental MGF candidate can be identifed with a
probability of 5 \%, and when excluding the Sagittarius galaxy after 6.4 years.

Combining the selection on $E_{\mathrm{iso}}$ with the duration $T_{90}$ and SFR
selection yields MGF candidates in {20.3 \%} of the surveys if the Sagittarius
Galaxy is included the computations, and 6.0 \% if the galaxy is excluded.
A detection of an sGRB fulfilling the spatial, $E_{\mathrm{iso}}$, $T_{90}$ and
SFR condition is detected with a 5 \% probability after 3.4 years of observation
time, and when excluding the Sagittarius galaxy after 11.7 years.

\subsection{Fluence Selection}
As a final selection, we use a threshold for the fluence of
$10^{-6} \, \mathrm{erg}\, \mathrm{cm}^{-2}$ to incorporate the IPN efficiency
at higher fluences of the source. When only considering the spatial and fluence
selection, GBM has a 5 \% detection probability after 2.6 years, and when
excluding the Sagittarius Galaxy after 7.1 years.

\begin{figure}[htpb]
  \centering
  \includegraphics[width=0.49\textwidth]{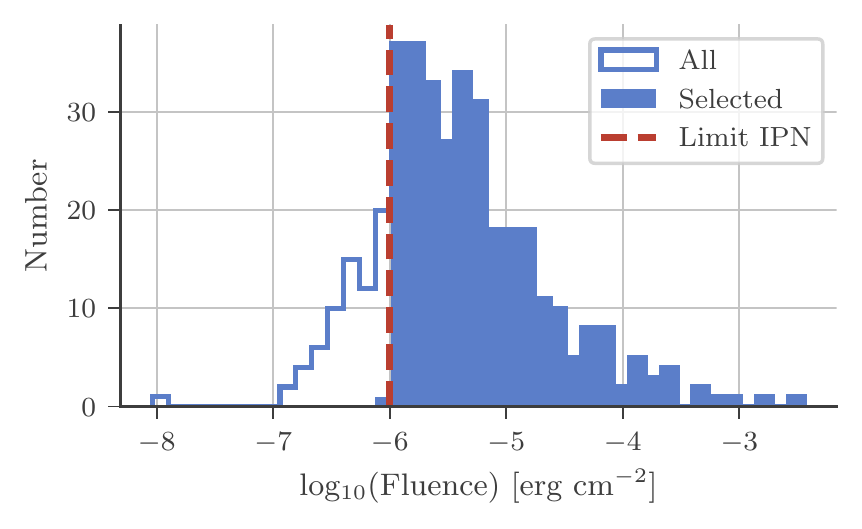}
  \caption{Histogram of the fluence of the simulated sGRBs that coincide with a
    nearby galaxy, and are selected by the GBM trigger algorithm.}
  \label{fig:selection_fluence}
\end{figure}

Finally, combining all proposed selections, namely the spatial, SFR, $T_{90}$,
$E_{\mathrm{iso}}$, and fluence selection, yields in 15.7 \% of the simulated
surveys coincidences in 9 different host galaxies.  When excluding the
Sagittarius Dwarf galaxy as host, there is only a 4.3 \% chance for chance
coincidences.  After an observation time of 16.3 years, there is a 5 \% chance
for coincidences, which is soon reached.

In Table \ref{tab:sum_results}, a summary of the results of all combinations of
selections is given.

\begin{table*}[htpb]
  \centering
  \caption[]{Summary of all results when applying the GBM Trigger, SFR,
    $T_{90}$, $E_{\mathrm{iso}}$, and fluence selections. The second and third
    column summarize the probability for misidentifying an sGRB as magnetar in an
    observation period of 14 years given the chosen selection criteria. The last
    two columns show the observation time with GBM that yields a 5 \% detection
    probability of sGRBs with the chosen characteristics.}
  \begin{tabular}{|c|c|c|c|c|}
    \hline
    \multicolumn{1}{|p{4.4cm}|}{\centering Selection} & 
                                                        \multicolumn{1}{p{2.7cm}|}{\centering Misidentification probability \\
    (all) [\%]} & \multicolumn{1}{p{2.7cm}|}{\centering Misidentification probability  \\  
    (without SagdSph) [\%]} & \multicolumn{1}{p{2.3cm}|}{\centering Observation time until 5\% probability \\
    (all) [yr]} & \multicolumn{1}{p{3.0cm}|}{\centering Observation time until 5\% probability \\ 
    (without SagdSph) [yr]}\\
    \hline
    \hline
    GBM & 41.1 & 15.3 & 1.7 & 4.6 \\
    GBM, SFR & 29.8 &  10.3 & 2.3 & 6.8 \\
    GBM, $T_{90}$ & 22.6 & 8.5 & 3.1 & 8.2 \\
    GBM, SFR, $T_{90}$ & 20.5 & 6.2 & 3.4 & 11.3 \\
    GBM, $E_{\mathrm{iso}}$ & 30.4 & 11.0 & 2.3 & 6.4 \\
    GBM, SFR, $E_{\mathrm{iso}}$ & 28.0 & 7.7 & 2.5 & 9.1 \\
    GBM, SFR, $E_{\mathrm{iso}}$, $T_{90}$ & 20.3 & 6.0 & 3.4 & 11.7 \\
    GBM, Fluence & 26.6 & 9.8 & 2.6 & 7.1 \\
    GBM, Fluence, $T_{90}$ & 17.7 & 6.5 & 4.0 & 10.8 \\
    GBM, Fluence, $E_{\mathrm{iso}}$ & 25.8 & 9.0 & 2.7 & 7.8 \\
    GBM, Fluence, SFR & 23.8 & 6.3 & 2.9 & 11.1 \\
    GBM, Fluence, $T_{90}$, $E_{\mathrm{iso}}$ & 17.2 & 6.0 & 4.1 & 11.7 \\
    GBM, Fluence, $T_{90}$, SFR & 15.9 & 4.5 & 4.4 & 15.6 \\
    GBM, Fluence, $E_{\mathrm{iso}}$, SFR & 23.5 & 6.0 & 3.0 & 11.7 \\
    \multicolumn{1}{|p{4.4cm}|}{\centering GBM, Fluence,$T_{90}$, $E_{\mathrm{iso}}$,SFR} & 15.7 & 4.3 & 4.5 & 16.3 \\
    \hline
  \end{tabular}
  \label{tab:sum_results}
\end{table*}

\section{Conclusions}
\label{sec:conclusions}

In summary, after an exhaustive list of selections are applied to our
simulations, namely the spatial, GBM trigger, GBM duration $T_{90}$, isotropic
energy release $E_{\mathrm{iso}}$, and fluence selection, the probability of
detecting an sGRB which could be potentially misidentified as an extragalactic
MGF is 15.7 \%. When excluding the dominating Sagittarius Galaxy, the probability
of detecting an sGRB that fulfills all six criteria reaches the 5 \% level after
an observation period of about 16 years, which is not far in the future. While
the analysis cannot prove that a specific one of the previously proposed MGFs
are sGRBs, it attempts to integrate over the probability of a misidentification
by simulating an alternative hypothesis. Thus, any significance of true
association must incorporate the possibility that the event is actually an sGRB.

The number of coincidences grows as a power function of the error circle size of
the sGRBs localization for error circles larger than approximately 1.8
arcmin. While improving the localization decreases the number of galaxies in the
field of view, we showed that this is not enough to distinguish MGFs from sGRBs.

Hence, for the unambiguous identification of extragalactic MGFs in the future,
it will be necessary to find other criteria. Such a criterion is for example the
pulsating tail in the measured light curve, as clearly detected for the galactic
MGFs.  The oscillations in the tail are not characteristic of sGRBs, and could
thus distinguish sGRBs from MGFs. However, when MGFs originate from larger
distances, it becomes less likely that the pulsating tail is detectable.

A direct proof of an association of an MGF with a LV galaxy would be given by
measuring the distance to the source, which can be derived from measuring the
redshift. Thus, fast follow up observations in other wavelengths, especially in
the optical, and determining the redshift is important to identify the
gamma-rays' source. It is indeed true that many of the properties assigned to
current candidate MGFs rely entirely on the host galaxy's distance which can
lead to circular reasoning.

We conclude that an association of $\gamma$-ray sources with LV galaxies does
not necessitate the progenitor to be an MGF as there is a non-negligible
probability of the progenitor being an sGRB at larger distance behind the LV
galaxy along the line of sight. sGRBs can reproduce the same typical
characteristics of the measured $\gamma$-ray sources that were associated with
magnetars. Indeed, similar considerations for other spatial and/or temporal
coincidence should be made in order to minimize the chances of false
associations. For example, nearly all distances to sGRBs have been determined by
assuming that a spatially projected nearby galaxy is the host and thus this
galaxy's redshift is used for the sGRB. Our study could be adapted to compute
how likely this host association is made by chance coincidence when, for
example, the sGRB's true host is a more distant and undetected galaxy, or
alternatively a much closer and low-mass, low surface-brightness undetected
galaxy.

All code required to reproduce this work is distributed freely
\footnote{\url{https://github.com/grburgess/grb_shader}}.

\begin{acknowledgements}
  We thank Bjoern Biltzinger and Aaron Tohuvavohu for discussions.
  The final writing of this paper was funded by the Federal Ministry of
  Education and Research (BMBF) and the Baden-Württemberg Ministry of Science
  as part of the Excellence Strategy of the German Federal and State
  Governments.
  JMB acknowledges support from the Alexander von Humboldt Foundation and the
  DFG-funded Collaborative Research Center 1258.
  
\end{acknowledgements}

\bibliographystyle{aa}
\bibliography{references}

\end{document}